\documentclass[aps,twocolumn,prc,amsmath,amssymb,groupedaddress,showpacs]{revtex4-1}

\usepackage{graphicx}
\usepackage{soul}
\usepackage{tikz}
\usepackage{mathtools}
\usepackage{xspace}
\usepackage{color}
\usepackage{float}
\usepackage[normalem]{ulem}

\newcommand{\fd}[1]{\ensuremath{\mathrm{d}#1\:}}
\newcommand \unit[1] {$\,$#1}
\newcommand{\hp}{g\mu_{\mathrm B} h_{\mathrm {p}}}
\newcommand{\hap}{g\mu_{\mathrm B} h_{\mathrm {ap}}}
\newcommand{\hop}{g_{\mathrm {orb}} \mu_{\mathrm{orb}} h_{\mathrm {p}}^{\mathrm{orb}}}
\newcommand{\hoap}{g_{\mathrm {orb}} \mu_{\mathrm{orb}} h_{\mathrm {ap}}^{\mathrm{orb}}}
\newcommand{\Gp}{\ensuremath{G_{\mathrm {p}}}}
\newcommand{\Gap}{\ensuremath{G_{\mathrm {ap}}}}
\newcommand{\Vg}{\ensuremath{V_{\mathrm {g}}}}

\DeclareMathOperator{\su}{\!\uparrow}
\DeclareMathOperator{\sd}{\!\downarrow}

\DeclareMathOperator{\Rep}{Re}
\DeclareMathOperator{\Imp}{Im}
\DeclareMathOperator{\Tr}{Tr}
\DeclareMathOperator{\im}{i\!}
\DeclareMathOperator{\Ha}{\hat H}

\DeclareMathOperator{\ro}{\hat \rho}

\DeclareMathOperator{\hc}{\,h.c.}

\DeclareMathOperator{\No}{\hat N}

\DeclareMathOperator{\dig}{\hat\Psi^{(0)}\!}

\DeclareMathOperator{\DSO}{\Delta_{\mathrm {SO}}}
\DeclareMathOperator{\kbt}{k_{\mathrm {B}}\! T}
\DeclareMathOperator{\mub}{\mu_{\mathrm {B}}\!}
\DeclareMathOperator{\morb}{g_{\mathrm{orb}}\mu_{\mathrm {orb}}\!}
\DeclareMathOperator{\Vb}{V_{\mathrm {b}}}

\usetikzlibrary{patterns}
\usetikzlibrary{shapes.geometric}
\usetikzlibrary{scopes}
\usetikzlibrary{trees}
\usetikzlibrary{decorations.pathmorphing}
\usetikzlibrary{decorations.markings}
\usetikzlibrary{calc}
\usetikzlibrary{chains}
\usetikzlibrary{positioning}
\usetikzlibrary{decorations}
\usetikzlibrary{decorations.pathreplacing}
\usetikzlibrary{snakes}

\tikzset{
    re/.style={postaction={decorate},
        decoration={markings,mark=at position .85 with {\arrow[scale=2]{>}}}},
    li/.style={postaction={decorate},
        decoration={markings,mark=at position .15 with {\arrow[scale=2]{<}}}},
    elf/.style={draw=blue, postaction={decorate},
        decoration={markings,mark=at position .55 with {\arrow[scale=2,draw=blue]{>}}}},
    elb/.style={draw=blue, postaction={decorate},
        decoration={markings,mark=at position .45 with {\arrow[scale=2,draw=blue]{<}}}},
      rm/.style={solid,->},
      lm/.style={dashed,->},
      refnod/.style={circle,draw,fill,scale=.5},
      pole/.style={circle,scale=.3,fill=blue,pin distance=2cm}
}

\begin{document}

\title{Transport across a carbon nanotube quantum dot 
contacted with ferromagnetic leads: experiment and non-perturbative modeling}

\author{Alois Dirnaichner}
\email[]{alois.dirnaichner@ur.de}
\author{Milena Grifoni}
\affiliation{Institute for Theoretical Physics, 
University of Regensburg, 93040 Regensburg, Germany}
\author{Andreas Pr\"ufling}
\author{Daniel Steininger}
\author{Andreas K. H\"uttel}
\author{Christoph Strunk}
\affiliation{Institute for Experimental and Applied Physics, 
University of Regensburg, 93040 Regensburg, Germany}

\date{\today}

\begin{abstract}
We present measurements of tunneling magneto-resistance (TMR) in
single-wall carbon nanotubes attached to ferromagnetic contacts in the
Coulomb blockade regime. Strong variations of the TMR with gate
voltage over a range of four conductance resonances, including a
peculiar double-dip signature, are observed. The data is compared to
calculations in the "dressed second order" (DSO) framework. In this
non-perturbative theory, conductance peak positions and linewidths are
affected by charge fluctuations incorporating the properties of the
carbon nanotube quantum dot and the ferromagnetic leads. The theory is
able to qualitatively reproduce the experimental data.
\end{abstract}

\pacs{73.23.Hk, 73.63.Fg, 75.76.+j, 72.25.-b, 85.75.-d}

\maketitle

\section{Introduction}
Controlling electronic spin in nano-scale circuits is a long-lasting
challenge on the way to fast-switching, energy-efficient building
blocks for electronic devices. To this end, spin-dependent transport
properties have been investigated in a wealth of low dimensional
systems, e.g., mesoscopic magnetic islands~\cite{Zaffalon2003},
2DEGs~\cite{Hu2001}, InAs nanowires~\cite{Hofstetter2010},
graphene~\cite{Tombros2007} and
fullerenes~\cite{Pasupathy2004}. Carbon nanotubes (CNTs), being thin,
durable and high-throughput wiring, allow coherent transport of
electronic charge and spin and are promising candidates for future
spintronics applications~\cite{Shulaker2013}.
While control and
scalability of CNT-based nanocircuits still pose significant
challenges, devices where single carbon nanotubes (CNTs) are contacted to ferromagnetic
leads can be produced with standard lithography methods:
spin valve experiments were performed on
single-wall~\cite{Sahoo2005,Kim2002,Jensen2005,Nagabhirava2006}
(SWCNT) and
multi-wall~\cite{Tsukagoshi1999,Orgassa2001,Alphenaar2001,Zhao2002,Chakraborty2003}
carbon nanotubes in various electron transport regimes. 
In most cases, a spatially confined quantum dot is
coupled to ferromagnetic electrodes. Electronic transport across CNT
quantum dots can take place in different regimes: Depending on the
relative magnitude of coupling strength, temperature and charging
energy, this ranges from an opaque Coulomb-blockade
regime~\cite{Rudzinski2001,Wetzels2005,Koller2007,Braun2004}, to an
intermediate coupling regime with lead induced energy level
shifts~\cite{Cottet2006,Koller2012,Kern2013}, to a strongly correlated
Kondo regime~\cite{Lopez2003,Martinek2003,Gaass2011,Schmid2015}. For
highly transparent contacts, in contrast, the dot behaves essentially like
an electronic wave guide~\cite{Cottet2006a,Man2006}.

In our work, we focus on the conductance of a carbon nanotube quantum 
dot weakly coupled to ferromagnetic contact electrodes, recorded for 
parallel ($\Gp$) and anti-parallel ($\Gap$) contact magnetization, 
respectively. $\Gp$ and $\Gap$ define the so-called tunneling 
magneto-resistance (TMR) \cite{Cottet2006a,Barnas2008}: 
$\mathrm{TMR}=(\Gp/\Gap)-1$. 
Experimentally, the TMR has been shown to be strongly gate 
dependent~\cite{Sahoo2005,Samm2014}. We report on shifting 
and broadening of conductance peaks resulting in specific dip-peak and dip-dip
sequences in the TMR gate dependence. Our data covers a range of four
Coulomb resonances with extremal TMR values of $-20\%$ to $+180\%$.

The pronounced resonant structure of the conductances \Gp\ and \Gap\
leads to large TMR values if the positions and widths of the
resonances depend on the magnetization configurations p and ap. 
Thus, various mechanisms
have been proposed which induce a shift of the energy levels of the
quantum dot, and thus the of the resonance peaks, 
depending on the magnetization of the
contacts. Those are spin-dependent interfacial phase
shifts~\cite{Sahoo2005} or virtual charge fluctuation
processes~\cite{Koller2012,Gaass2011}. The effect of spin polarized leads on the
resonance width have been described in~\cite{Tolea2006} for a resonant
single level junction. Interestingly, a negative TMR is predicted for
asymmetric couplings to the leads. An attempt to account for
broadening in the presence of Coulomb interactions was discussed
within a self-consistent approach based on the 
equation of motion (EOM) technique~\cite{Cottet2006}. 
The EOM was applied to model TMR data
reported for a SWCNT~\cite{Sahoo2005} for a model with spin-dependent 
interfacial phase shifts.

Here we discuss a transport theory which naturally
incorporates the effects of spin polarized
leads on the position and width of conductance resonances in the
presence of strong Coulomb interactions. It is an extension of
the so-called dressed second order (DSO) transport theory, recently
developed for normal leads~\cite{Kern2013}, to the case of
spin-polarized contacts. This theory accounts for energy
renormalization and broadening of the peaks in linear conductance due
to charge fluctuation processes. We show that the charge fluctuations also
affect transport through excited states in the non-linear conductance
regime. This observation is in agreement with previous reports on
tilted co-tunneling lines in CNT quantum dots~\cite{Holm2008}. A
qualitative agreement with the experimental findings is obtained.

This paper is structured as follows. We first present the measurement
details and experimental data in Sec.~\ref{sec:exp}. In
Sec.~\ref{sec:theory} we introduce the so-called dressed second order
theory (DSO)~\cite{Kern2013} in the reduced density matrix transport
framework and address its implications on non-linear conductance and
TMR. Finally, in Sec.~\ref{sec:comp} we provide a comparison between
experimental data and results from the DSO and draw our conclusions in
Sec.~\ref{sec:sum}.
 
\section{Experiment}
\label{sec:exp}
\subsection{Sample preparation}

For the purpose of measuring TMR in CNTs, one needs to interface the
nanotube to two ferromagnetic contacts with a different switching
field. The conductance, being sensitive to the magnetization in the
leads, changes when the polarization of one of the contacts is
reversed by an external magnetic field. It has been shown that
NiFe is well suited as a material for the electrodes of CNT
spin-valves~\cite{Aurich2010}: the alloy shows a distinct switching 
behavior as a
function of the applied magnetic field and the interface transparency 
between NiFe and the CNT is comparable to that of Pd.
\begin{figure}
\centering
\includegraphics[width=.48\textwidth]{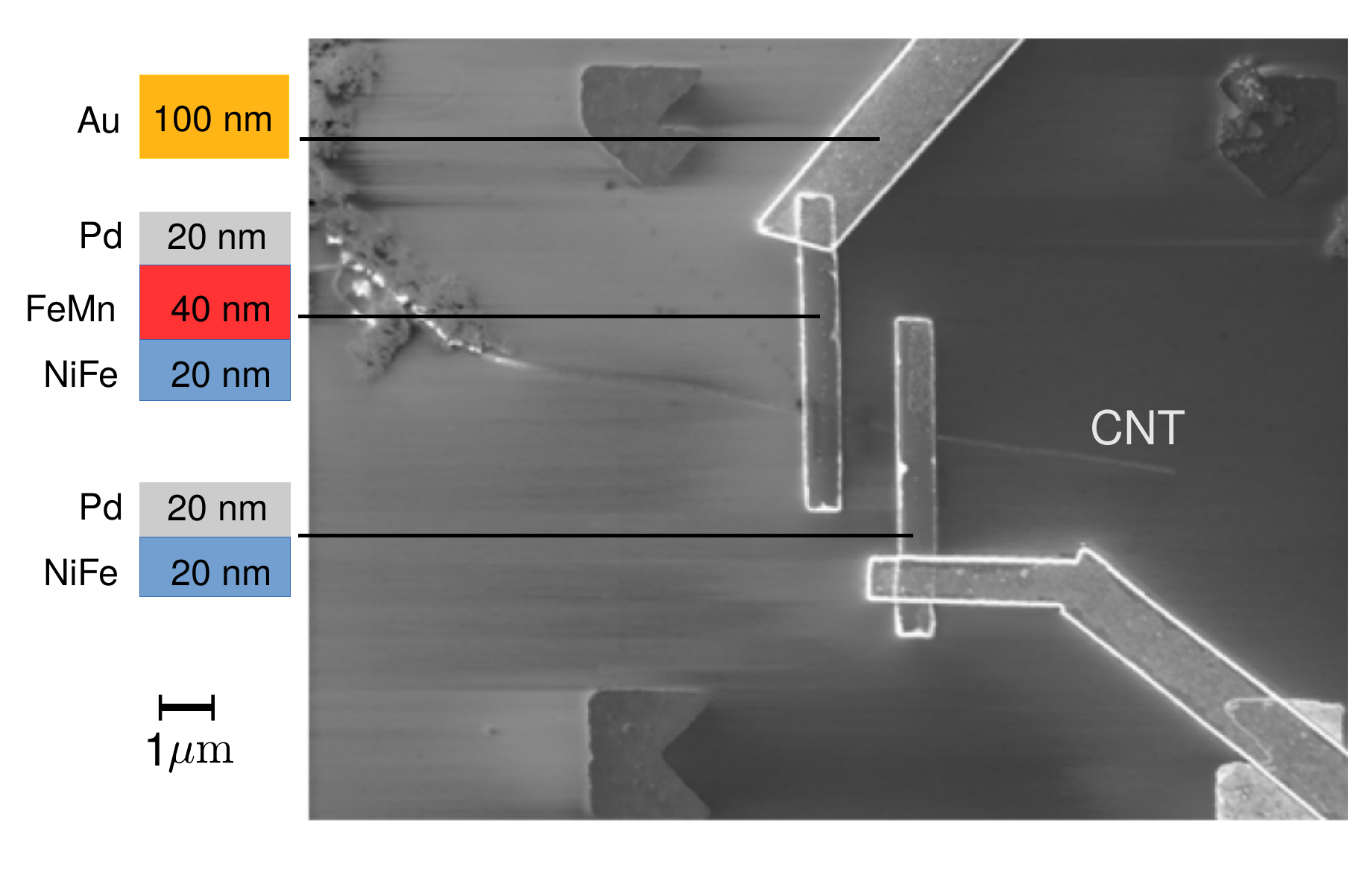}
\caption{(Color online) SEM picture of a chip structure similar to that of the
measured device. A carbon nanotube on a positively doped silicon
substrate capped with $500$\unit{nm} $\text{SiO}_2$ is contacted by
two Permalloy stripes, one of which is exchange-biased by a FeMn
layer. On top, the stripes are protected by palladium. Gold is used
for the bond pads and the connections to the nanotube
contacts.\label{sample}}
\end{figure}  
The structure of one of the devices we realized for this purpose is
shown in Fig.~\ref{sample}. On an oxidized silicon substrate
($500$\unit{nm} $\text{SiO}_2$) a carbon nanotube is grown by chemical
vapor deposition. The nanotube is located by atomic force microscopy
and two NiFe (80:20) leads, $20$\unit{nm} in thickness, are deposited
at a distance of $1$\unit{$\mu$m} on top of the nanotube by
sputtering. On one of the two contacts, $40$\unit{nm} of
anti-ferromagnetic FeMn (50:50) is sputtered to bias the magnetization
of the underlying NiFe contact. The hysteresis loop of this contact is
expected to be shifted with respect to the pure NiFe
contact by virtue of the exchange bias effect~\cite{Choo2007}.
A $20$\unit{nm} protective layer (Pd) covers
the leads from the top.
The switching of the exchange biased contacts was confirmed
independently prior to the measurement using SQUID and vibrating
sample magnetometer techniques. 

\subsection{Measurement}
\label{sec:meas}

\begin{figure}
\centering
\includegraphics[width=.48\textwidth]{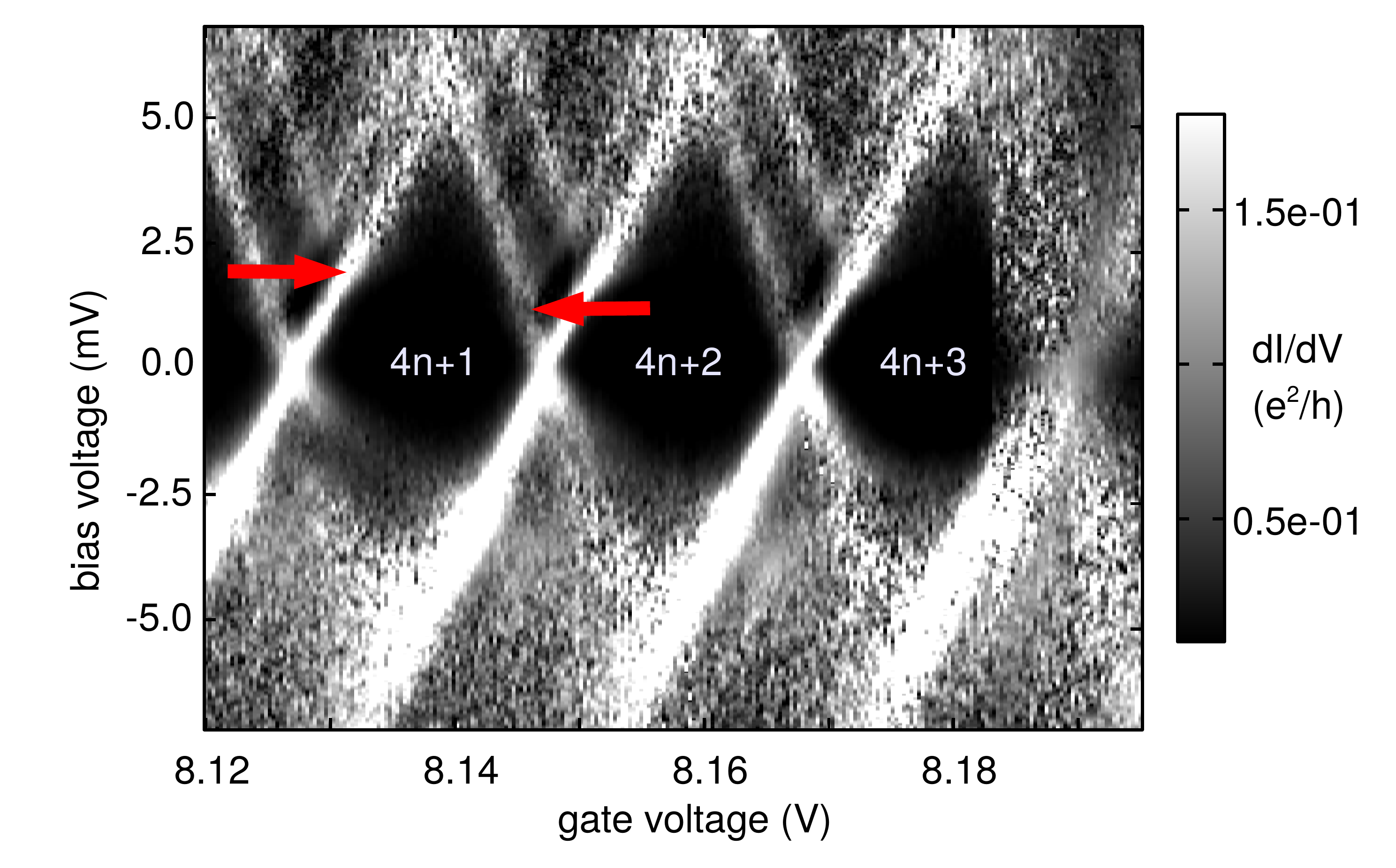}
\caption{(Color online) Differential conductance versus bias and gate voltage of a
selected region measured at $300$\unit{mK} and $B=0$. The numbers in the Coulomb
blockade regions denote the number of electrons in shell $n$ on the
quantum dot. Arrows indicate the first excited state crossing the
source (left) and drain (right) lines in the vicinity of the state
with one extra electron ($N=4n+1$).\label{cd}}
\end{figure}  
An electronic characterization of the quantum dot at $300$\unit{mK}
and at zero magnetic field
shows a regular Coulomb blockade behavior (Fig.~\ref{cd}). The data
yield a gate conversion factor $\alpha = 0.29$ and a charging energy
of $E_{\mathrm c} = 6.1$\unit{meV} (see Eq.~(\ref{eq:hcnt})). The
sample does not exhibit a clear four-fold symmetry in the peak height
or peak spacing as expected for a carbon nanotube quantum
dot. Consequently, we are not able to label the Coulomb blockade
regions with a value of the electronic shell filling $n$ in a definite
way. The assignment of the number of electrons to the experimental
data in Fig.~\ref{cd} is done in agreement with the theoretical
predictions in Sec.~\ref{sec:tmr}.

Having a closer look at Fig.~\ref{cd}, we can identify an excited
state transition at $1.4$\unit{meV} parallel to the source line (left
arrow) and at $\sim 1.8$\unit{meV} parallel to the drain line (right
arrow). The energy scale of this excitation stays approximately
constant over a range of at least six resonances, as can be seen from
measurements over a broader gate range. The quantization energy
$\epsilon(n)$ of a CNT shell $n$ is a direct consequence of the
electron confinement along the nanotube. It yields a mean level
spacing $\epsilon_0=\epsilon(n+1)-\epsilon(n)\propto\hbar v_{\mathrm
F}/\pi L$, where $L$ is the CNT length. It is thus reasonable to
identify the first excitation with the confinement energy $\epsilon_0$
equivalent to a lateral confinement of $1.1$\unit{$\mu$m} for a Fermi
velocity of $800$\unit{km/s}~\cite{Dresselhaus1996}, a value close to
the contact spacing of $1$\unit{$\mu$m}. The asymmetry of the line
spacing with respect to source and drain suggests a gate-dependent
renormalization~\cite{Holm2008} of the CNT many-body addition energies
in the presence of ferromagnetic contacts. We show in
Sec.~\ref{sec:exc_state_ren} that this can be a direct consequence of
charge fluctuations in the presence of contact magnetization.

\begin{figure}
\centering
\includegraphics[width=.48\textwidth]{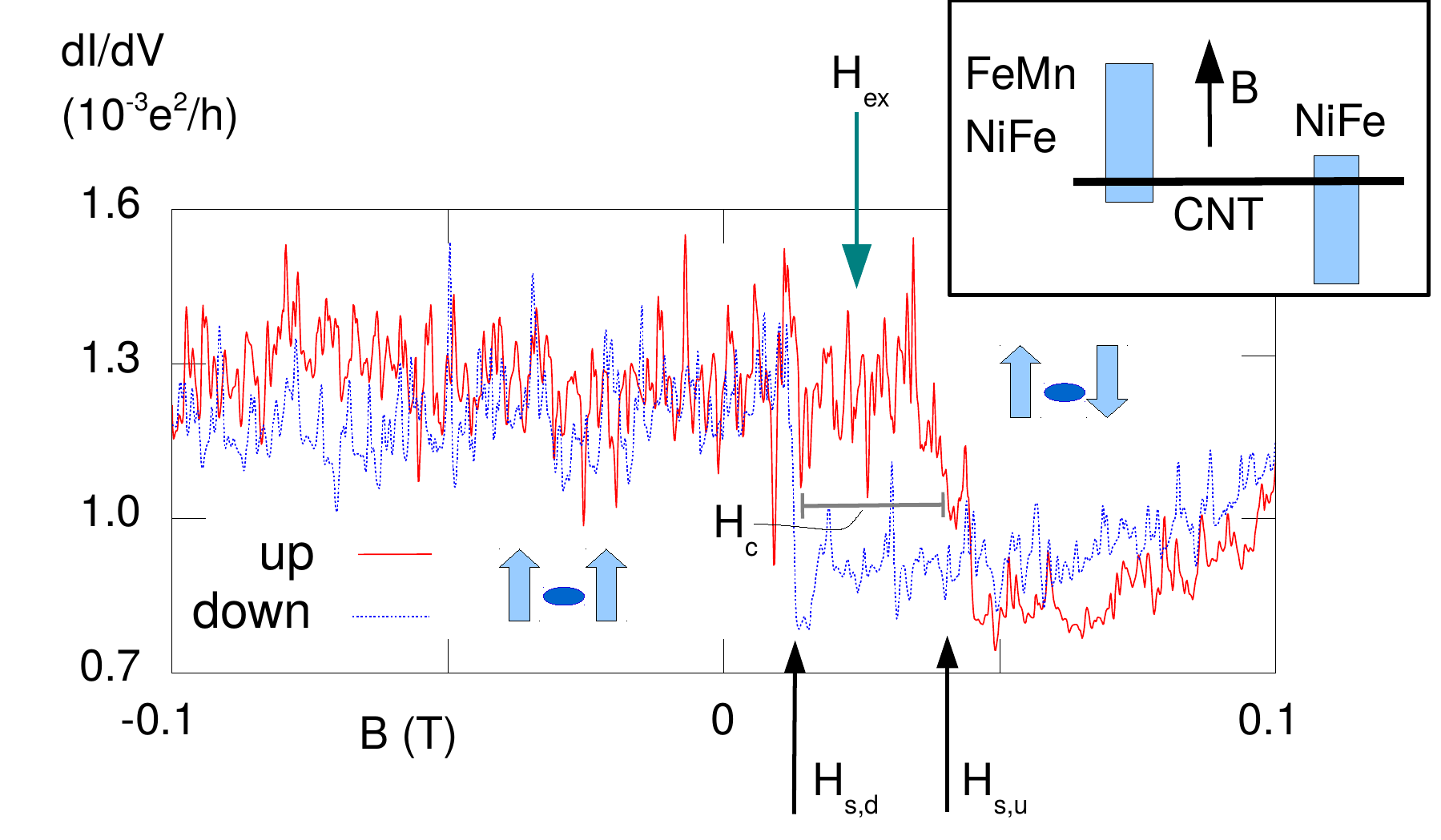}
\caption{(Color online) Differential conductance plotted versus
magnetic field at $\Vg=8.1737$\unit{V}, $\Vb=0$ and $300$\unit{mK}. The solid
red curve was recorded with increasing field, the dashed blue curve
with decreasing field. Small pictograms indicate possible orientations
of the majority spins in the contacts. The switching of one of the two
contacts at $H_{\mathrm{s,u/d}}$ is highlighted with
arrows at the bottom for both sweep directions. The coercive field is
indicated by $H_{\mathrm c}$ and the exchange bias by $H_{\mathrm
  {ex}}$. 
Inset: Orientation of
the external field $B$ with respect to the CNT and the
leads.\label{fig:switch}}
\end{figure}

Electron transport measurements at
$300$\unit{mK} show a significant switching behavior. In
Fig.~\ref{fig:switch} the conductance across the CNT quantum dot is
plotted against the magnetic field directed parallel to
the stripes, i.e., along their easy axis, as indicated in the inset to
the figure. The steps in the signal can be interpreted as the
magnetization reversal of the contacts, as sketched in the figure. 
Sweeping the magnetic field from negative
($-100$\unit{mT}) to positive values, one of the contacts switches at
$H=H_{\mathrm{s,u}}$, resulting in a configuration with anti-parallel
polarization of the majority spins of the two contacts. This results in a
drop of the conductance signal. Upon increasing the field further, the
second contact is supposed to switch and the conductance should
recover. The second switching event was not observed in the present
sample. Sweeping back from positive to negative field, the
  conductance recovers at $H_{\mathrm{s,d}}$. 
The two values $H_{\mathrm{s,d/u}}$ characterize
  a hysteresis loop with a coercive field $H_{\mathrm
    c}=H_{\mathrm{s,u}}-H_{\mathrm{s,d}}$ and an exchange bias
  $H_{\mathrm {ex}}=(H_{\mathrm{s,u}}+H_{\mathrm{s,d}})/2$. 
At $B=0$ the two contacts are always in a parallel 
configuration, because the coercive field of the
switching contact is smaller than the exchange bias. 

\begin{figure}
\centering
\includegraphics[width=.48\textwidth]{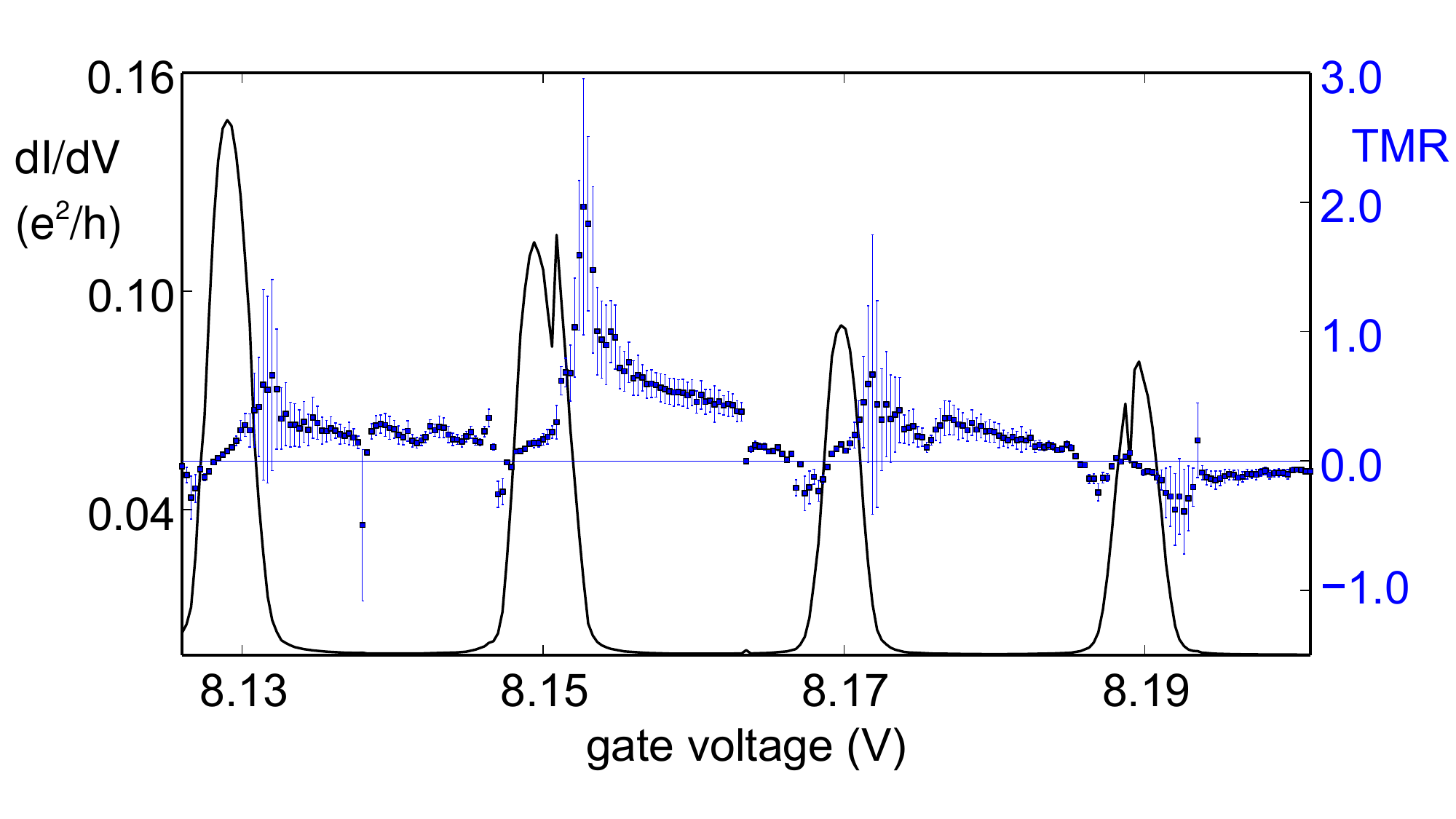}
\caption{(Color online) Differential conductance and TMR as a function of gate
voltage measured over four resonances (\textit{slow} measurement, see text). 
The conductance is
measured at parallel polarization of the contacts. The TMR graph shows
a dip-peak sequence over the first two resonances and a qualitatively
different double-dip feature at the last two.\label{tmr3dia}}
\end{figure}

Measurements
of the conductance performed at zero magnetic field require 
$\Delta t^{\mathrm {fast}} \sim 100$\unit{ms} per data point 
and will be called the \textit{fast} measurements in the
following. 
Contrarily, in \textit{slow} measurements, each conductance data
point is obtained from magnetic field sweeps with a duration
of $\Delta t^{\mathrm{slow}}\sim 20$ minutes at constant gate
voltage (compare Fig.~\ref{fig:switch}). We then identify $H_{\mathrm
  s}$ from a step in the conductance signal and take the average over 100 points
on either side of the step to extract the conductance in the parallel
and anti-parallel configuration, respectively. 
This is repeated for 250 values of the backgate potential
in the range between $8.126$\unit{V} and
$8.201$\unit{V}.
In Fig.~\ref{tmr3dia}, the TMR as a function of gate voltage is
shown together with the conductance at parallel contact polarization.  
In this \textit{slow} measurement, we obtain conductance peaks 
with a height of $0.15e^2/h$ and a full width at half maximum (FWHM) 
of $\Gamma\sim 0.7$\unit{meV}. Comparing these values to a height of
$0.3e^2/h$ and a width of $0.4$\unit{meV}
obtained from the \textit{fast} measurement at $B=0$ we conclude that
the peak conductance in the data from the $slow$ measurement is 
substantially suppressed.
We will discuss this deviation in Sec.~\ref{sec:comp}. It is remarkable that
besides huge positive (180\%) TMR values, negative regions occur prior
to the peak in the TMR curve in the first two resonances while for the
last two the value drops again, forming two dips in sequence. Again
this will be discussed in more detail in Sec.~\ref{sec:comp}.

\section{Theoretical modeling}
\label{sec:theory}

We proceed by presenting a theoretical framework capable to
reproduce the transport data from the previous section. In particular, the
connection between the theory and the resulting shape of the TMR curve
will be discussed in detail. 
In order to be able to account for a gate
dependence of the TMR, the transport theory should be able to
incorporate the influence of the ferromagnetically polarized leads on
the positions of the linear conductance maxima as well as on the width
of the conductance peaks. 

Noticeably, the commonly used perturbative
description of the Coulomb resonances predicts temperature broadened
peaks and maxima whose positions are solely determined by the isolated
quantum dot spectrum implying a constant, positive
TMR~\cite{Koller2012}.

\begin{figure}
\centering
\includegraphics[width=.48\textwidth]{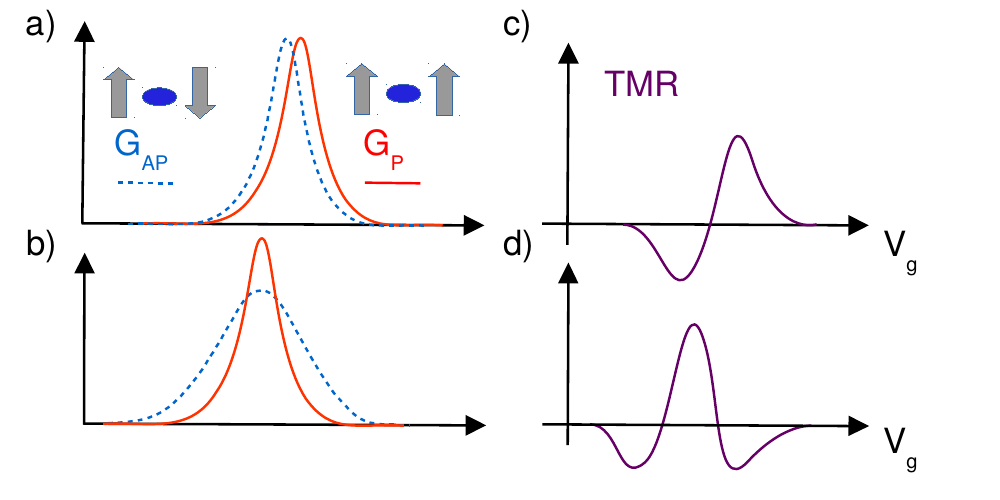}
\caption{(Color online) Left panels: Schematic drawing of the lead induced,
polarization dependent, modification of position (a) and width (b) of
a peak in the conductance across a quantum dot as a function of the
gate voltage. Right panels: As a consequence of the level shift (a)
and level broadening (b), the corresponding TMR signal exhibits a
characteristic dip-peak (c), or dip-peak-dip (d)
feature.\label{fig:intro}}
\end{figure}  
A transport theory accounting for charge fluctuations non-perturbatively
was shown to shift the quantum dot energy levels depending on
the magnetization configuration of the leads~\cite{Koller2012}. The
qualitative effect of the renormalization is depicted in
Fig.~\ref{fig:intro}(a): The peak in the conductance $\Gp$ in presence
of leads with parallel spin polarization is shifted with respect to
the one in $\Gap$, the conductance in the anti-parallel case. This shift
yields a characteristic dip-peak feature in the TMR signal, similar to
what was observed in Ref.~\citenum{Sahoo2005}. Yet, this theory cannot
account for the double-dip like TMR signatures visible in our data (see
Fig.~\ref{tmr3dia}, $\Vg\sim 8.19$\unit{V} and $\Vg\sim 8.17$\unit{V}). 
These require additionally a change of the resonance line-width when 
switching from the parallel to the anti-parallel configuration, as
shown in Fig.~\ref{fig:intro}(b) and 
also observed lately in Ref.~\cite{Samm2014}.

In the following, we discuss how to theoretically account for
broadening and renormalization effects, to lowest order in the
coupling $\Gamma$, within the recently proposed ``dressed second order
approximation'' (DSO). The DSO has been discussed in
Ref.~\citenum{Kern2013} for the single impurity Anderson model with
normal metal leads, where it has been shown to
correctly capture the crossover from thermally broadened to tunneling
broadened conductance peaks. Here we present its
generalization to a multilevel system coupled to ferromagnetic leads.

\subsection{Hamiltonian}
\label{sec:hamiltonian}
We treat the system as an isolated quantum dot coupled to metallic
leads. The Hamiltonian of such a system reads 
$\Ha=\Ha_{\mathrm R}+\Ha_{\mathrm D}+\Ha_{\mathrm T}$. Here,
\[ \Ha_{\mathrm R}=\sum_{l\sigma \mathbf k}\epsilon_{l\sigma \mathbf
k}\hat{c}^{\dagger}_{l\sigma \mathbf k}\hat{c}_{l\sigma \mathbf k}
\] is the Hamiltonian of an ensemble of non-interacting electrons in
the leads $l=\mathrm{s/d}$ with wave vector $\mathbf k$ and spin
$\sigma$. The operator $\hat{c}_{l\sigma \mathbf k}$
($\hat{c}^{\dagger}_{l\tau\sigma k}$) annihilates (creates) an
electron with energy $\epsilon_{l\sigma \mathbf k}$. The second part,
\begin{align}\label{eq:hcnt} \Ha_{\mathrm{D}}&=\frac 12 E_c \No^2 +
\sum_{n\tau\sigma}\left[\epsilon(n)+\tau\sigma \frac{\Delta_{\mathrm
{SO}}(n)}{2}\right]\No_{n\tau \sigma}\nonumber\\ &\quad-e\alpha
V_{\mathrm g}\No+\Ha^{\mathrm{P/A}}_{\mathrm {ext}},
\end{align} describes the electrons on the CNT quantum dot in terms of
the quantum numbers $n$ (shell), spin $\sigma$ and valley $\tau$. Here
we used
$\No_{n\tau\sigma}=\hat{d}^\dagger_{n\tau\sigma}\hat{d}_{n\tau\sigma}$,
with the fermionic dot operator $\hat{d}_{n\tau\sigma}$ and
$\No=\sum_{n\tau\sigma}\No_{n\tau\sigma}$, the total dot
occupation. For our purposes, it is sufficient to account for Coulomb
interaction effects in terms of a capacitive charging energy
$E_{\mathrm c}$. Short range exchange contributions are neglected
here. The symbols $\tau$ and $\sigma$ represent the eigenvalues $\pm
1$ of the states with quantum numbers $K,K'$ and $\su,\sd$,
respectively. In the CNT, a non-zero spin-orbit coupling $\DSO$ can
lead to the formation of degenerate Kramer
pairs~\cite{Schmid2015}. Notice that, for simplicity, a valley mixing
contribution is not included in Eq.~(\ref{eq:hcnt}), as it would not
affect the main conclusions drawn in this work. Hence, the valley
degree of freedom is a good quantum number to classify the CNT's
states~\cite{Marganska2014}. The next to last part of the Hamiltonian
$\Ha_{\mathrm D}$ models the effect of an electrostatic gate voltage
$V_{\mathrm g}$ scaled by a conversion factor $\alpha$. Finally,
$\Ha^{\mathrm{p/a}}_{\mathrm {ext}}$ accounts for external influences
on the dot potential, e.g., stray fields from the contacts and the
external magnetic field used to switch the contact polarization.
\begin{table}
\begin{tabular}{l l l} \toprule 
$N_{\mathrm {\mathrm{rel}}}$ 
& $\Delta_{\mathrm{SO}}\leq \max\{\kbt,\gamma_0\}$ 
& $\Delta_{\mathrm{SO}}\gg \max\{\kbt,\gamma_0\}$\\ \colrule
-1
&\parbox[t]{2.9cm}{\raggedright
$|K\su,K\sd,K'\su;n-1\rangle$\\
$|K\su,K\sd,K'\sd;n-1\rangle$\\ 
$|K\su,K'\su,K'\sd;n-1\rangle$\\
$|K\sd,K'\su,K'\sd,n-1\rangle$} 
&\parbox[t]{2.9cm}{\raggedright
$|K\su,K\sd,K'\su;n-1\rangle$\\
$|K\su,K\sd,K'\sd,n-1\rangle$}\\ 
0
&$|n\rangle$
&$|n\rangle$ \\
1
&\parbox[t]{2.2cm}{\raggedright
$|K\su;n\rangle \; |K\sd;n\rangle$\\
$|K'\su;n\rangle \; |K'\sd;n\rangle$}
& $|K\su;n\rangle \;|K\sd;n\rangle$\\ 
2
&\parbox[t]{4cm}{\raggedright
$|K\su,K\sd;n\rangle\; |K\su,K'\su;n\rangle$\\ 
$|K\su,K'\sd;n\rangle \;|K\sd,K'\su;n\rangle$\\ 
$|K\sd,K'\sd;n\rangle \;|K'\su,K'\sd;n\rangle$}
& $|K\su,K\sd,n\rangle $\\
3
&\parbox[t]{2.4cm}{\raggedright
$|K\su,K\sd,K'\su;n\rangle$\\
$|K\su,K\sd,K'\su;n\rangle$\\ 
$|K\su,K'\su,K'\sd;n\rangle$\\
$|K\sd,K'\su,K'\sd;n\rangle$}
&\parbox[t]{2.4cm}{$\raggedright
|K\su,K\sd,K'\su;n\rangle$\\
$|K\su,K\sd,K'\sd;n\rangle$} \\ 
4
&$|n+1\rangle$
&$|n+1\rangle$ \\
5
&\parbox[t]{3.2cm}{\raggedright
$|K\su;n+1\rangle \;|K\sd;n+1\rangle$\\ 
$|K'\su;n+1\rangle \; |K'\sd;n+1\rangle$}
&$|K\su;n+1\rangle \; |K\sd;n+1\rangle$ \\ 
\botrule
\end{tabular}
\caption{The set of allowed electronic ground states $C$ of the CNT
with $N$ electrons for large (right) and small (left) spin-orbit
coupling $\Delta_{\mathrm{SO}}$. The degeneracy of the configuration
depends on the magnitude of $\DSO$. In the first column, the excess
electron number $N_{\mathrm {rel}}=N-4n$ is reported with respect to
the number $4n$ of electrons in the filled $(n-1)$-th
shell.\label{tab:eigenstates}}
\end{table}

The ground states of shell $n$ have $4n+a$ ($0\leq a\leq 3$) electrons
and will in the following be characterized by the quantum numbers of
the excess electrons with respect to the highest filled shell
$n-1$. For instance, the quantum dot state labeled by
$|K\su;n\rangle$ contains $4n$ electrons plus one additional electron
in the ($K,\su$) state. Including states with $4n-1$ and $4n+5$
electrons we end up with 6 ground states with different degeneracies
(see Tab.~\ref{tab:eigenstates}, left column). In total we consider a
Fock space of dimension $24$ if the four-fold degeneracy is not lifted
by a sufficiently large spin-orbit coupling $\Delta_{\mathrm
{SO}}$. The extra states with occupation $4n-1$ and $4n+5$ are
included to allow for charge fluctuations in and out of the shell $n$
under consideration. Conversely, for large enough spin-orbit coupling
the dimension of the Fock space is reduced to 10, see
Tab.~\ref{tab:eigenstates}, right column. Judging from the stability
diagram in Fig.~\ref{cd} and from data over a greater gate range where
we see no two-fold pattern in the spacing of the excited state lines,
we consider the configuration on the left side in
Tab.~\ref{tab:eigenstates} to be more likely. For a compact notation,
the shell number will in the following be neglected from the state ket
if not necessary.

Quantum dot and metallic leads are coupled perturbatively by a
tunneling Hamiltonian
\begin{align}\label{eq:HT} \Ha_{\mathrm T}=
\sum_{l \mathbf k n\sigma\tau}T_{l \mathbf k n\sigma\tau}d^{\dagger}_{n\sigma\tau}
c_{l \mathbf k\sigma} + \hc,
\end{align} with a tunnel coupling $T_{l\mathbf k n\sigma\tau}$
generally dependent on the quantum numbers of both leads and quantum
dot. In the following, for simplicity, we assume that $T_{l\mathbf
\mathbf k n\sigma\tau}=T_{l}$.

\subsection{The reduced density matrix within 
the dressed second order (DSO) approximation}
\label{sec:diagram}

We describe the state of our system by the reduced density matrix
$\ro=\Tr_{\mathrm {R}}\{\ro_{\mathrm {tot}}\}$, obtained by tracing
over the possible configurations of states in the reservoirs, assuming
that they are in thermal equilibrium. For the quantum dot itself we
suppose that it reaches a steady state characterized by
$\dot{\ro}=0$. The corresponding stationary Liouville equation
reads~\cite{Kern2013}
\begin{align}\label{eq:liouville}
0 &= -\im\sum_{aa'}\delta_{ab}\delta_{a'b'}(E_a-E_a')\rho_{aa'}+
\sum_{aa'}K^{aa'}_{bb'}\rho_{aa'},
\end{align}  
in terms of matrix elements $\rho_{ab}=\langle a|\hat{\rho}|b\rangle$
of $\hat \rho$ in the eigenbasis of the quantum dot. The superoperator
$K$ connects initial states $|a\rangle$, $|a'\rangle$ to final states
$|b\rangle$ and $|b'\rangle$ at a certain order in the perturbation
$\Ha_{\mathrm T}$.

The calculation of the kernel elements is performed along the lines of
Ref.~\citenum{Kern2013}. As an example, the element connecting the
states $|b\rangle$, $|b'\rangle = |b\rangle$ and $|a\rangle$,
$|a'\rangle = |a\rangle$ is given in second order by
\[
K^{aa}_{bb}=\sum_l\Gamma^p_{l,ba}=\sum_l\frac \im\hbar \lim_{\lambda\to 0^+} 
\int\fd{\epsilon}\frac{\gamma_{l}^{ba}(\epsilon)f^{p}_l(\epsilon)}
{E_{a}^{b}-\epsilon+\im\lambda}+\hc,
\]
where $\Gamma^{p}_{l,ba}$ is the corresponding tunneling rate. The
function $f^p_l(\epsilon)$ with $p=\pm$ is defined as
$f^\pm_l(\epsilon)=[1+\exp\{\pm\beta(\epsilon-\mu_l)\}]^{-1}$, where
$\beta$ is the inverse temperature and $\mu_l$ the lead's chemical
potential. Hence, $f^+_l(\epsilon)=f_l(\epsilon)$ is the Fermi
function and describes the occupation probability in lead $l$. In
general, $p=\pm 1$ if the final state $|b\rangle$ has one electron
more/less than the initial state $|a\rangle$.
The energy difference between final and initial dot configuration is
given by $E^b_a = E_b-E_a=\tilde E_b - \tilde E_a-e\alpha
\Vg(N_b-N_a)$. Finally,
\[ 
\gamma^{ba}_{l}(\epsilon) =
\gamma_{l\sigma(b,a)}(\epsilon)=|T_l|^2{\cal D}_{l\sigma}(\epsilon)
\] 
is a spin-dependent linewidth defined in terms of the tunneling
amplitude $T_l$ and of the spin-dependent density of states ${\cal
D}_{l\sigma}(\epsilon)$. A Lorentzian provides a cut-off for the
density of states at a bandwidth $W$. The notation $\sigma(a,b)$
indicates that the spin $\sigma$ of the electron tunneling out of/onto
lead $l$ depends on the spin configuration of the initial state $a$
and the final state $b$ of the quantum dot. It is convenient to
introduce the spin-resolved density of states of lead $l$ at the Fermi energy
\begin{equation}\label{eq:polarization}
{\cal D}_{l\sigma}={\cal D}_{l\sigma}(\epsilon_{\mathrm F})=D_0(1+\sigma P_l)/2
\end{equation}
where $P_l=({\cal D}_{l\uparrow}-{\cal D}_{l\downarrow})/({\cal
D}_{l\uparrow}+{\cal D}_{l\downarrow})$ is the polarization of lead
$l$. The couplings $|T_l|^2$ we define in the same spirit as
\begin{equation}\label{eq:coupling_asym}
|T_{s/d}|^2=|T_0|^2(1\pm a)/2,
\end{equation}
using the parameter $a$ to tune the asymmetry in the coupling to the leads.
We will in the following use the factorization
\begin{equation}\label{eq:kappa}
\gamma_{l\sigma(b,a)}(\epsilon_{\mathrm F})=\gamma_0\kappa_{l\sigma},
\end{equation}
where we collect the lead and spin independent prefactors in an
overall coupling strength $\gamma_0=D_0|T_0|^2$ and include the
dependence on spin and lead index in the dimensionless parameter
$\kappa_{l\sigma}$, where $\sum_{l\tau\sigma}\kappa_{l\sigma}=1$. Note
that $\gamma_0$ is related to the level broadening $\Gamma_0$ by
$\Gamma_0 = 2\pi\gamma_0$.

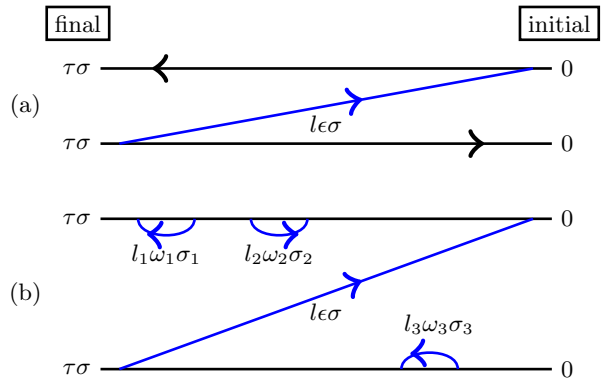
\begin{figure}
\vspace*{.5cm}
\centering
\begin{tikzpicture}
\tikzstyle{every path}=[line width=1pt]
\draw (-.3,4.6) node[draw] {final};
\draw (6.1,4.6) node[draw] {initial};
\coordinate (b1) at (0,0);
\coordinate (b2) at (0,2);
\coordinate (a1) at (0,3);
\coordinate (a2) at (0,4);
\draw (b1) node[left]{$\tau\sigma$} -- +(6,0) node[right]{$0$};
\draw (b2) node[left]{$\tau\sigma$} -- +(6,0) node[right]{$0$};
\draw[elb] ($(b2)+(5.75,0)$) -- 
node[midway,right,below]{$l\epsilon\sigma$}($(b1)+(.25,0)$);
\draw[elb] ($(b2)+(0.5,0)$) to[out=-90,in=-90] 
node[midway,below]{$l_1\omega_1\sigma_1$} +(.75,0);
\draw[elb] ($(b2)+(2.75,0)$)  to[out=-90,in=-90] 
node[midway,below]{$l_2\omega_2\sigma_2$} +(-.75,0);
\draw[elb] ($(b1)+(4,0)$) to[out=90,in=90] +(.75,0); 
\draw ($(b1)+(4.5,.6)$) node {$l_3\omega_3\sigma_3$};
\draw[re] (a1) node[left]{$\tau\sigma$} -- +(6,0) node[right]{$0$};
\draw[li] (a2) node[left]{$\tau\sigma$} -- +(6,0) node[right]{$0$};
\draw[elb] ($(a2)+(5.75,0)$) -- 
node[midway,right,below]{$l\epsilon\sigma$} ($(a1)+(.25,0)$);
\draw (a1)+(-1,.5) node{(a)} (b1)+(-1,1) node{(b)};
\end{tikzpicture}
\caption{(Color online) Diagrammatic representations of the
contributions to the rate $\Gamma^{+}_{l,\tau\sigma0}$ in second order
(a), and an example of diagrams included in the DSO (b). In the latter
case, the fermion line (blue) from the second order theory is
``dressed'' by charge fluctuation processes. The labels below the
fermion lines denote energy and spin of the particle tunneling
from/onto the lead. Note that the diagram is read from right to left,
i.e., the initial state $|0\rangle$ can be found on the right and the
final state $|\tau\sigma\rangle$ on the left.\label{fig:diagrams}}
\end{figure}

In Fig.~\ref{fig:diagrams}(a), a diagrammatic representation of one
contribution to the second order kernel is shown for the case of
$|a\rangle = |0\rangle$ and $|b\rangle=|\tau\sigma\rangle$. The
fermionic line connecting the lower to the upper contour carries
indices $l,\epsilon,\sigma$ which fully characterize the nature of the
electron tunneling between lead $l$ and quantum dot. The direction of
the arrow further specifies if the electron tunnels out of (towards
lower contour) or onto (towards upper contour) the dot.

\begin{figure}
\centering
\includegraphics[width=.48\textwidth]{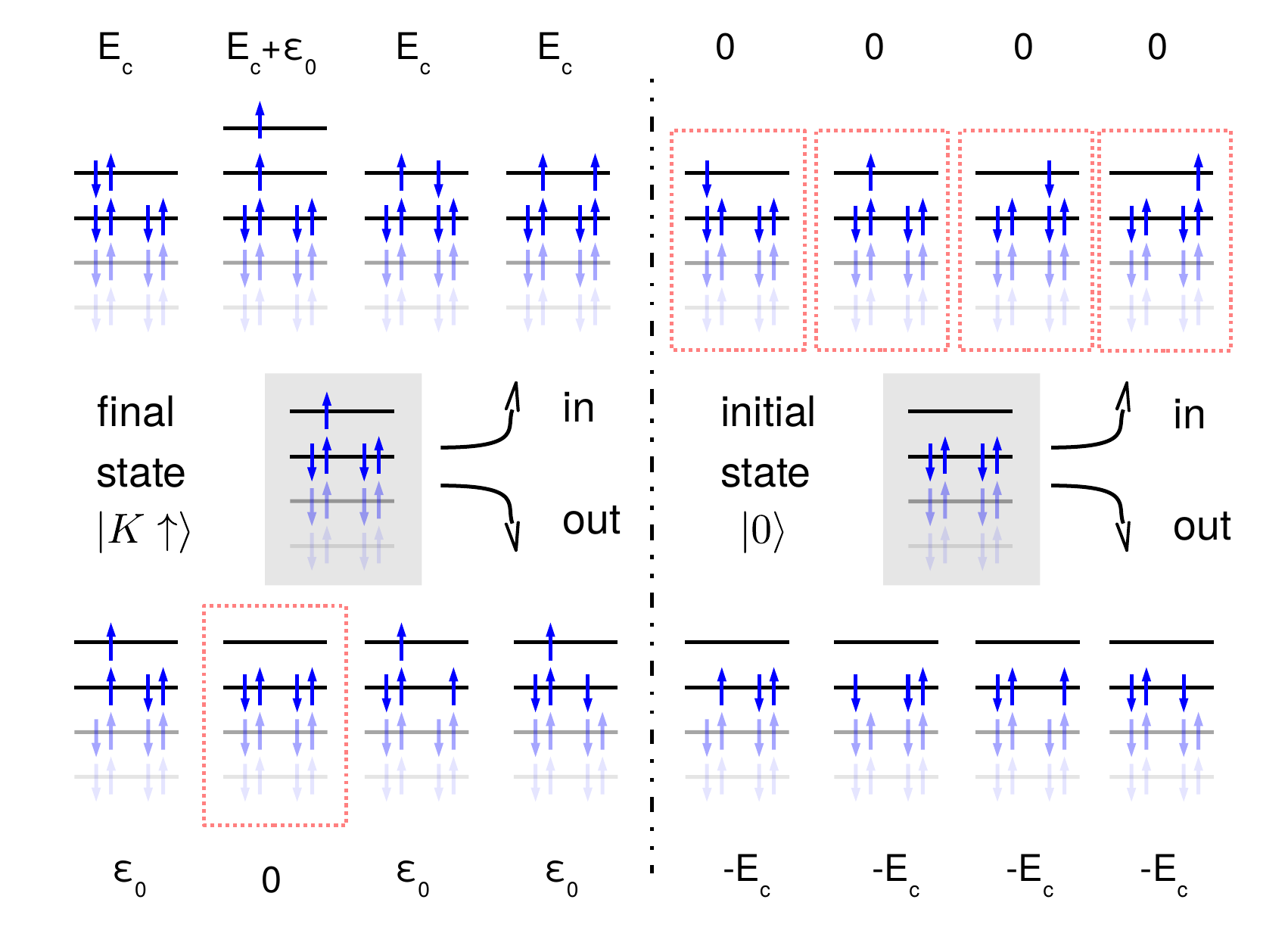}
\caption{Example of possible charge fluctuations for a final state
($|K\su\rangle$, left, shaded gray) with one extra electron and an
initial state with zero electrons in the shell $n$ ($|0\rangle$,
right, shaded gray). This set corresponds to one specific diagram of
the type shown in Fig.~\ref{fig:diagrams}(b). States that can be
reached by in-tunneling of an electron are shown on top, states that
can be reached by out-tunneling of an electron are shown on the
bottom. Dashed frames highlight resonant ($E^{c'/b}_{a/c'}=0$) charge
fluctuations. Above and below the level schemes, the energy difference
between the virtual state and the state on the other contour is given:
the energies of the states accessible from the initial (final) state
are compared to the energy of the final (initial) state on resonance
($\tilde E^{K\uparrow}_0=e\alpha \Vg$). Note that the electron number
of the states that can be reached by in-tunneling on the left and the
number of electrons in the initial state on the right differ by
two. The same situation occurs for the final state and the
out-tunneling states on the right. The energy differences for this
class of fluctuations is of the order of $E_{\mathrm c}$. A comparison
of the electron number of the final state with the in-tunneling states
on the left and the initial state with the out-tunneling states on the
right yields a difference of zero. These fluctuations have comparably
low energy cost. \label{tmr_argument}}
\end{figure}

Beside this lowest (second) order contribution, we consider all
diagrams of the structure shown in Fig.~\ref{fig:diagrams}(b). The
selected diagrams contain arbitrary numbers of uncorrelated charge
fluctuation processes (bubbles in Fig.~\ref{fig:diagrams}). During the
charge fluctuation, the dot state on the upper contour has one charge
less or more compared to that of the final state
$|\tau\sigma\rangle$. Hence, the virtual state is either the state
$|0\rangle$ or one of the many (see Tab.~\ref{tab:eigenstates}) doubly
occupied states. On the lower contour, the fluctuations take place
with respect to the initial state $|0\rangle$. Examples of charge
fluctuations in the case of initial state $|0\rangle$ and final state
$|K\su\rangle$ are shown in Fig.~\ref{tmr_argument}. Summing all
diagrams of this type yields the DSO rates :
\begin{align}\label{rateq}
\Gamma^{+}_{l,ba}&= \frac 1{2\pi\hbar} 
\int\fd{\epsilon}\nu^{ba}_l(\epsilon)f^+_l(\epsilon),\nonumber\\
\end{align}
for a state $b$ that can be reached by an in-tunneling process from state $a$, and
\begin{align}
\Gamma^{-}_{l,ab}&= \frac 1{2\pi\hbar} \int\fd{\epsilon}\nu^{ba}_l(\epsilon)f^-_l(\epsilon)
\end{align}  
for an out-tunneling process $b\to a$. 
Note that we introduced a tunneling-like density of states (TDOS)
\begin{align}\label{eq:tdos}
\nu^{ba}_l(\epsilon)= \frac{\gamma_l^{ba}(\epsilon)\Imp(\Sigma^{ba}(\epsilon))}
{[\Imp(\Sigma^{ba}(\epsilon))]^2+[\epsilon-E^b_a+\Rep(\Sigma^{ba}(\epsilon))]^2}.
\end{align}
We refer to the contribution $\Sigma^{ba}$ in the denominator of the
TDOS as a self energy that infers from the contributions of all
possible charge fluctuations connected to the initial, $a$, and final,
$b$, states in the state space given in
Tab.~\ref{tab:eigenstates}. Explicitly,
\begin{align}\label{eq:selfenergy}
\Sigma^{ba}(\epsilon)&=\sum_{\substack{c\in\{b,a\}\\c'\in C^{\pm}_c}}a_{ba}^{c'c}(\epsilon),
\end{align}
with the sets $C_{b/a}^{\pm}$ given by
\begin{align}
C_{b/a}^{\pm}\coloneqq\{c' : N_{c'}=N_{b/a}\pm 1 \wedge 4n-1\leq N_{c'}\leq 4n+5\}.
\end{align}
The sets are shown in Fig.~\ref{tmr_argument} for the states
$|a\rangle = |0\rangle$ and $|b\rangle=|K\uparrow\rangle$.

The summand
\[
a_{ba}^{c'(b/a)}(\epsilon)=\sum_l \int\fd{\omega}
\frac{\gamma^{c'(b/a)}_l(\omega)f^{p}_l(\omega)}
{\pm p\omega+\epsilon-E^{c'/b}_{a/c'}+\im\eta}
\]
accounts for a transition from $b$ or $a$ to a state $c'$, with $c'\in
C_{b/a}^{p}$.  Performing the integral, we arrive at an analytic
expression for the contributions to the self energy, i.e.,
\begin{widetext}
\begin{align}\label{eq:as}
a_{ba}^{c'(b/a)}(\epsilon)&=
\sum_l\gamma^{c'(b/a)}_l(\epsilon)
\left\{\im\pi f^p_l(\pm p(E^{c'/b}_{a/c'} -\epsilon))\pm \left[\dig\left(W\right)
-\Rep\left[\dig\left(\im\, (\mu_l\pm p(E^{c'/b}_{a/c'} -\epsilon))\right)\right]
\right]\right\},
\end{align}
\end{widetext}
where $\dig(x)=\Psi^{(0)}(0.5+x/2\pi\kbt)$ and $\Psi^{(0)}$ is the
digamma function. Note that the dependency on the bandwidth drops out
due to the alternating sign of the contributions from the upper and
lower contour in the summation in Eq.~(\ref{eq:selfenergy}).  Having
calculated the self energy, we are now able to collect all rates
according to the transitions in our state space, and solve the
stationary Eq.~(\ref{eq:liouville}) to obtain the occupation
probabilities $\rho_{aa} = P_a$. Within the steady state limit we can
neglect off-diagonal entries $\rho_{ba}$ if they are among
non-degenerate states~\cite{Koller2012}. According to
Tab.~\ref{tab:eigenstates}, the CNT spectrum can be spin and valley
degenerate. However, the tunneling Hamiltonian (\ref{eq:HT}) conserves
the spin during tunneling, and thus spin coherences are not present in
the dynamics. Here, for simplicity, orbital coherences are neglected
as well \footnote{For CNTs of the zig-zag type, coherences are not
expected to contribute to the dynamics for tunneling processes which
conserve the crystal angular momentum, i.e., for which the
perpendicular component $k_\perp$ of the momentum $\mathbf k$ is
conserved during tunneling. This is because in zig-zag type CNTs the
two valleys correspond to different values of the crystal angular
momentum.}.

\subsection{Current within the DSO}
\label{sec:conductance}
The current through the terminal $l$ can be written in terms of the
difference of in- and out-tunneling contributions at the
junction~\cite{Alhassid2004}:
\begin{align}\label{eq:current}
&I_l(\Vb)=\frac {e}{2\pi\hbar}\times\nonumber\\
& \sum_{\substack{a \in C\\ c \in C^{+}_a}} \int \fd{\epsilon}
\left[ P_a(\Vb)f^+_l(\epsilon) - P_c(\Vb)f^-_{l}(\epsilon) \right] 
\nu_l^{ca}(\epsilon,\Vb),
\end{align}
where $\Vb$ is the bias voltage applied between the two contacts, and
$C$ is the set of all possible configurations (see
Tab.~\ref{tab:eigenstates}). In general, the populations can be
expressed in terms of rates via the Liouville equation
(\ref{eq:liouville}) and a closed form for the current and,
consequently, for the conductance can be found. This is
straightforward if two states are connected by pairwise gain-loss
relations~\cite{Alhassid2004}. For the case of the single impurity
Anderson model, for example, a compact notation of the conductance can
be given~\cite{Kern2013}. In this work, the conductance data from 
the model is calculated numerically~\footnote{The source code for the
  numerical calculation can be found on
  https://github.com/Loisel/tmr3.}.

The width of a resonance in conductance with respect to the gate
potential is determined by the populations, the TDOS which has a form
similar to a Lorentzian, and by the derivative of the Fermi
functions. Note that the populations are themselves a function of the
rates and therefore are also governed by the resonance conditions of
the rates. The DSO theory has been proven to be quantitatively valid
down to temperatures $4\kbt \sim \gamma_0$ in the single electron
transistor \cite{Kern2013}. Upon decreasing of the temperature below
$\gamma_0/4$, a quantitative description of the transition rate
$\Gamma^{ac}_l$ would require to calculate $\Sigma$ beyond the lowest
order in $\gamma_0$.  In the regime where temperature and coupling are
of comparable magnitude, the width and position of the Coulomb
blockade peaks in a gate trace are strongly influenced by the TDOS
and, more precisely, by the self energy $\Sigma$. The role of
$\Rep(\Sigma)$ is to influence the positions of the Coulomb blockade
peaks: In the rate for the transition $a$ to $b$, the real part
appears next to the energy difference $E^b_a$ of the transition in the
denominator. Hence, due to this contribution the resonant level is
shifted depending on the configuration of the leads.

\subsection{Renormalization of excited states}
\label{sec:exc_state_ren}

\begin{figure} \centering
\includegraphics[width=\columnwidth]{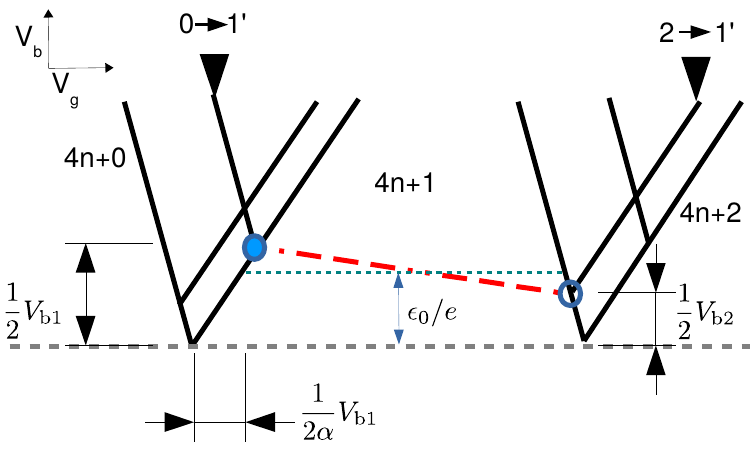}
\caption{Schematic drawing of the conductance lines in the vicinity of
the charging state with $4n+1$ electrons in Fig.~\ref{cd}. The first
visible excitation is shifted upwards on the left and downwards on the
right side of one charging diamond by $e(V_{\mathrm b1}-V_{\mathrm
b2})/2=-\delta_1$. The corresponding energies in Fig.~\ref{cd} are
$eV_{\mathrm b1}/2\simeq 2$\unit{meV} and $eV_{\mathrm b2}/2\simeq
1.4$\unit{meV}. For our analysis we choose bias and gate voltages
close to the filled dot for the first transition $0\to 1'$ and to the
empty circle for the second transition $2\to 1'$. \label{exc_state}}
\end{figure}

In the stability diagram in Fig.~\ref{cd} we observe an asymmetry in
the spacing of lines associated with excited states connected to one
charging state, as drawn schematically in Fig.~\ref{exc_state}. The
line $0\to 1'$ meets the diamond at bias voltage $V_{\mathrm
b1}$. Measured along the bias voltage axis, this value is larger than
the energy difference $V_{\mathrm b2}$ associated with the line $2\to
1'$ on the right. A similar behavior has been discussed previously for
the co-tunneling regime~\cite{Holm2008}. As noted by these authors,
the asymmetry can not be explained within the sequential tunneling
picture but can be attributed to the renormalization of the excitation
energies $E^b_a$ in Eq.~(\ref{eq:tdos}) due to virtual tunneling
processes. Although the framework in Ref.~\cite{Holm2008} is
different, the evaluation of $\Rep(\Sigma^{ba})$ is similar to that in
our model. The condition for a resonance for a transition between
states $a$ and $b$ is given by
\begin{align}\label{eq:res_cond} \epsilon \pm e\!\Vb/2 + e\alpha\!\Vg
- \tilde E^b_a + \Rep(\Sigma^{ba})=0,
\end{align} 
where $\epsilon$ is the energy of the tunneling electron
with respect to the chemical potential of the unbiased contact
$\mu_0$.  Note that this condition can be fulfilled for different
transitions at the same time, a situation that occurs at any point
where two lines in a stability diagram intersect.  In order to
interpret the observed shift of the excited state line in the
differential conductance data in Fig.~\ref{tmr_argument}, it is
illuminating to study the contribution from $\Rep(\Sigma)$ at points
$(V_{\mathrm{g1}},V_{\mathrm{b1}})$ and
$(V_{\mathrm{g2}},V_{\mathrm{b2}})$ marked by a dot and a circle,
respectively, in Fig.~\ref{exc_state}.
We consider an exemplary set of states $0=|0;n\rangle$,
$1=|K\su;n\rangle$, $1'_1=|[K\su];n\rangle$,
$1'_2=|K\su,K\sd,(K'\su);n\rangle$ and $2=|K\su,K\sd;n\rangle$. A
similar analysis can be carried out for other states with $4n+1$ and
$4n+2$ electrons. The quantum numbers in round brackets denote a
missing electron of shell $n-1$ whereas the square brackets indicate a
state of shell $n+1$. For each of the highlighted points in
Fig.~\ref{exc_state}, two conditions in the form of
Eq.~(\ref{eq:res_cond}) can be given. Subtracting them pairwise we are
left with
\begin{align}\label{eq:res_cond_1}
e\! V_{\mathrm b1}-E^{1'}_1+[\Rep(\Sigma^{1',0})-\Rep(\Sigma^{1,0})]&=0,\\
\label{eq:res_cond_2}
e\! V_{\mathrm b2}-E^{1'}_1+[\Rep(\Sigma^{2,1})-\Rep(\Sigma^{2,1'})]&=0,
\end{align}
where the self energy contributions depend on bias and gate
voltage. To lowest order in $\gamma_0$ we analyze the differences in
$\Rep(\Sigma)$ using $e\!\Vb_{1/2}=E^{1'}_1$ and $\alpha
e\!\Vg^{1/2}=\tilde E^{1/2}_{0/1}\pm e\!\Vb/2$ at $\epsilon = 0$. In
order to calculate $\Rep(\Sigma)$ we have to analyze the contributions
from all accessible states in Eq.~(\ref{eq:selfenergy}). In principle
there are arbitrarily many states that can be reached by a charge
fluctuation. However, we assert that the available energy interval for
charge fluctuation processes is given by
$\max(e\!\Vb,\Gamma_0,3-4\kbt)$ and contributions beyond this scale
are suppressed. Numerical results using a larger bandwidth can be
found in Sec.~\ref{sec:cfb} of the appendix.

For our considerations we assume that the spin orbit coupling of our
CNT quantum dot is small, i.e., $\DSO<\max(\kbt,\Gamma)$. Otherwise we
would expect to see a two-fold symmetry in the spacing of the excited
state lines in the stability diagram in Fig.~\ref{cd}. The other
important scales - charging energy, shell spacing and linewidth - are
related in the way $E_{\mathrm c}>\epsilon_0\gg\max(\kbt,\gamma_0)$. 
Within this choice of parameters the difference of the self 
energy corrections for the resonant
transition can be calculated
by~(\ref{eq:res_cond_1})$-$(\ref{eq:res_cond_2})$=0$, i.e.,
\begin{align}\label{eq:delta1}
&\delta_{1}\equiv\left[\Rep(\Sigma^{1'0})-\Rep(\Sigma^{10})\right] - 
\left[\Rep(\Sigma^{21})-\Rep(\Sigma^{21'})\right]\nonumber\\
&\quad\simeq \gamma_0\left\{-1 + 2\bar\kappa_{\mathrm s}\!
-\!\bar\kappa_{\mathrm d}+\bar\kappa_\uparrow\!
-\!\bar\kappa_\downarrow\right\}\Psi^0_{\mathrm R}(\epsilon_0/2)
\end{align}
where we used the abbreviation $\Psi^0_{\mathrm
R}(\epsilon)=\Rep[\Psi^0(1/2+\im\epsilon/2\pi\kbt)]$ and a bar denotes
a summation over indices, e.g., $\bar\kappa_l=\sum_\sigma
\kappa_{l\sigma}$. A detailed derivation of these quantities is given
in the appendix, Sec.~\ref{sec:resigma}.  Similar calculations are
performed for the excited states in the $n+2$ and $n+3$ diamonds,
yielding
\begin{align*}
\delta_{2}&\simeq\gamma_0\left\{\bar\kappa_{\mathrm s}\!-\!\bar\kappa_{\mathrm{d}} 
+ \kappa_{\mathrm{s}\downarrow}\!-\!\kappa_{\mathrm{d}\uparrow} 
\right\}\Psi^0_{\mathrm R}(\epsilon_0/2),\\
\delta_{3}&\simeq\gamma_0\left\{1 + \bar\kappa_{\mathrm s}\!-\!2\bar\kappa_{\mathrm{d}} 
+\bar\kappa_\downarrow\!-\!\bar\kappa_\uparrow\right\}\Psi^0_{\mathrm R}(\epsilon_0/2),
\end{align*}
where the states with three electrons are chosen to be electron-hole
symmetric with respect to the state with one electron. 
Note that for the case of symmetric couplings the shifts reflect the 
electron-hole symmetry of the system while a choice of $a\neq 0$ 
(Eq.~(\ref{eq:coupling_asym})) breaks this symmetry. 
For highly asymmetric couplings
$|a|\sim 1$ the shifts are comparable to those in
Ref.~\cite{Holm2008}. Note that the effective change of the resonance
with respect to the energy difference has a negative sign (compare
Eq.~(\ref{eq:res_cond})). The resonance marked by the left arrow in
Fig.~\ref{cd} is situated above the resonance marked by the right
arrow. The experimental data thus corresponds to a negative shift. We
therefore assume an asymmetric coupling to the leads with a dominant
coupling to the drain contact, i.e., $\kappa_{\mathrm
s}<\kappa_{\mathrm d}$, $-1<a<0$. Using the parameters from a fit to
the data in Sec.~\ref{sec:sum}, i.e., $a=-0.7$ and
$\epsilon_0=1.4$\unit{meV} we obtain $\delta_1\approx -0.2$\unit{meV}
and $\delta_2 \approx -0.1$\unit{meV}. Compared to the shifts in the
experimental data, these values are too small by a factor of 2-3. We
expect that additional states may contribute to the charge
fluctuations that are not considered within this approximation.
  
\subsection{Tunneling magneto-resistance}
\label{sec:tmr}

Corrections to the conductance peak width are given by
$\Imp(\Sigma)$. Because $\Rep(\Sigma)$ and $\Imp(\Sigma)$ both depend
on the different magnetic properties of the source and drain leads as
well as on the dot's configuration, the resulting impact on the TMR is
quite intricate.  Thus we analyze the contributions to the self energy
in the light of different configuration of the lead's
polarizations. We focus on the last resonance, i.e., the transitions
$|0,n+1\rangle \rightleftarrows \{|(\sigma\tau),n+1\rangle\}$ where
the TMR graph in Fig.~\ref{tmr3dia} exhibits a double dip like
structure. The back-gate voltage is tuned such that
\[
\epsilon+e\alpha V_{\mathrm g}-\tilde E^0_{(\tau\sigma)}+\Rep(\Sigma^{0,(\tau\sigma)})=0,
\]  
and the quantum numbers in round brackets $(\tau\sigma)$ denote a
missing electron of shell $n+1$. At lowest order in the tunnel
coupling $\gamma_0$ we approximate $e\alpha V_{\mathrm g}=\tilde
E^0_{(\tau\sigma)}$ when we calculate
$\Rep(\Sigma^{0,(\tau\sigma)})$. From Eq.~(\ref{eq:as}) we list the
imaginary part of the self energy for this transition, i.e.,
\begin{align*}
&\Imp(\Sigma^{0,(\tau\sigma)})=\pi\gamma_0\sum_l\\
&\Bigg\{\sum_{c\in C^+_{0}}\!\kappa_{l\sigma(c)}f^+_l(E^{c}_{(\tau\sigma)}-\epsilon)
+\sum_{c'\in C^-_{0}}\!\kappa_{l\sigma(c')}f^-_l(\epsilon-E_{(\tau\sigma)}^{c'})\\
&+\!\!\sum_{c\in C^+_{(\tau\sigma)}}\!\!\!\kappa_{l\sigma(c)}f^+_l(\epsilon-E^{0}_{c})
+\sum_{c'\in C^-_{(\tau\sigma)}}\!\!\!\kappa_{l\sigma(c')}f^-_l(E_{c'}^{0}-\epsilon)\Bigg\}.
\end{align*}
The magnitude of the energy difference of the virtual state with
respect to the state on the other contour determines whether a
possible charge fluctuation contributes to the renormalization of the
self energy or not: a contribution $~f^+_l(E_{\mathrm c}-\epsilon)$,
e.g., is exponentially suppressed in the vicinity of the resonance.

\begin{figure}
\centering
\includegraphics[width=.48\textwidth]{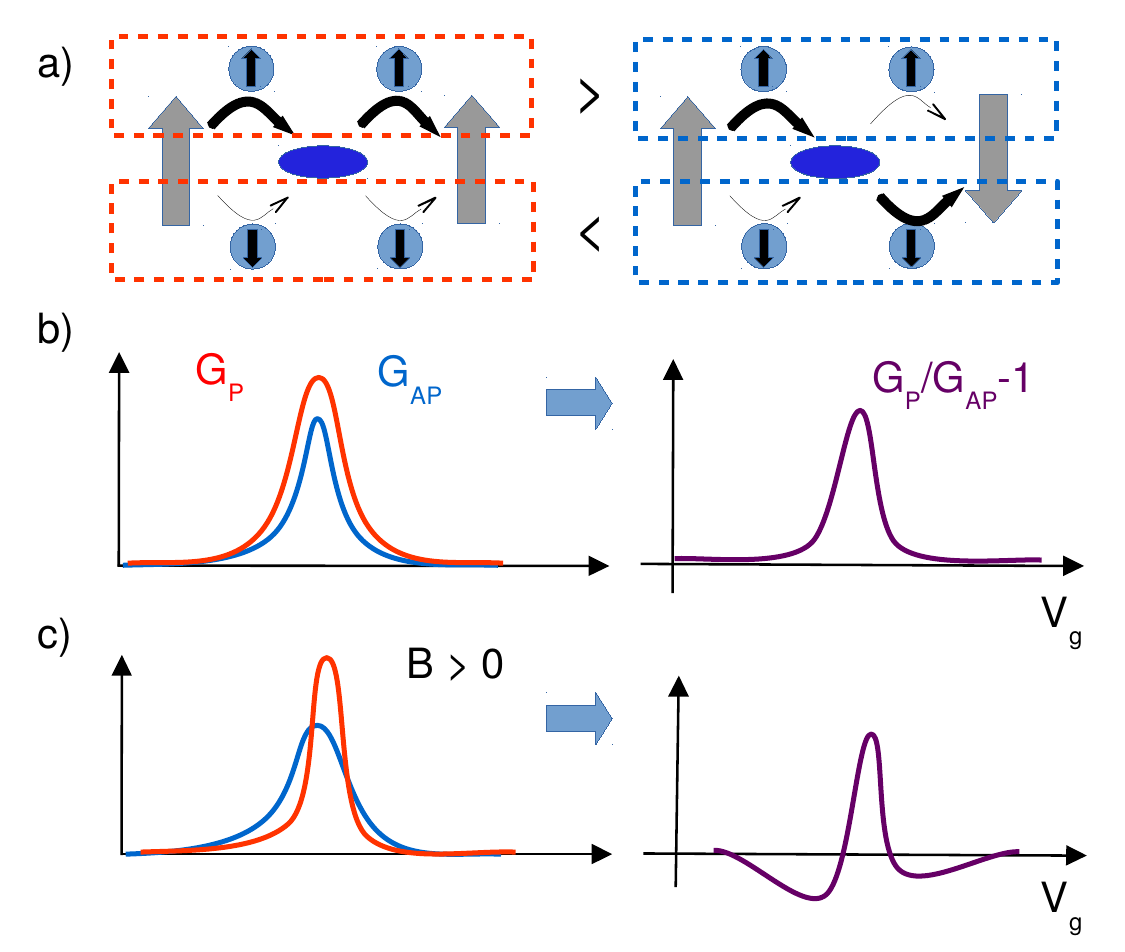}
\caption{(Color online) The influence of $\Imp(\Sigma)$ on the TMR. a)
Large gray arrows symbolize the majority spin in the left or right
contact. The contributions to the self energy for one spin species are
summed for each configuration of polarized leads (parallel on the
left, anti-parallel on the right) as indicated by the dashed
frames. Weak (strong) coupling to the dot (blue ellipse) is given by
thin (thick) arrows. Note that for the spin down species the sum over
the leads yields a greater contribution in the configuration with
anti-parallel polarization (as indicated by the signs between the
dashed frames). b) On the left, we depict schematically the
conductance peaks for one resonance in both parallel and anti-parallel
configurations and the resulting TMR (right). The broadening of $\Gp$
is typically larger than for $\Gap$ in the absence of stray fields. c)
Due to a magnetic stray field, the contribution to $\Imp(\Sigma)$ in
the parallel case can be reduced, giving rise to a double dip
structure in the TMR.\label{tmr_spin}}
\end{figure}

Therefore, knowing the arguments in the step functions $f^{\pm}$, we
can simplify the result significantly. Close to the resonance where
$|\epsilon| < \max(\kbt,\gamma_0)$, the fluctuations with an energy
cost of the charging energy $E_{\mathrm c}$ or of the shell spacing
$\epsilon_0$, e.g., the states that can be reached by out-tunneling
from the state $|(\tau\sigma)\rangle$ can be neglected. Focusing on
the resonant contributions, we are left with
\begin{align}\label{eq:Imdif}
&\frac{\Imp(\Sigma^{0,(\tau\sigma)})}{\pi\gamma_0}
\simeq\sum_l \left\{\vphantom{\sum_l}\kappa_{l\sigma}f^+_l(\epsilon)
+\sum_{\tau'\sigma'}\kappa_{l\sigma'}f^-_l(\epsilon-E^{(\sigma')}_{(\sigma)})\right\}.
\end{align}
It is clear from this result that the broadening of the TDOS peak does
depend on the lead configuration $\{\kappa_{l\sigma}\}$. Let the
majority spins be polarized such that $\sigma=+1$ in the layout with
parallel lead polarization. The sum over the leads is then given by
$\sum_{l}\kappa_{l\sigma}^{\mathrm p}=(1+\sigma P)/4$ and
$\sum_{l}\kappa_{l\sigma}^{\mathrm {ap}}=(1+\sigma P a)/4$ for
parallel and anti-parallel polarizations, respectively. Let us first
consider the case of zero effective Zeeman splitting, i.e.,
$E^{\sigma}_{\bar\sigma}=E^{(\bar\sigma)}_{(\sigma)}=0$. The
difference of $\Imp(\Sigma)$ for the two configurations then reads
\begin{align}\label{eq:Dims}
&\Imp\left[\Sigma^{0,(\tau\sigma)}_{\mathrm p}-\Sigma^{0,(\tau\sigma)}_{\mathrm {ap}}\right]=
\delta^{\Imp}=\pi\frac {\gamma_0}4 \sigma P (1-a) f^+(\epsilon).
\end{align}
Note that the validity of this result depends on the ratio of
linewidth and level spacing, namely that $\gamma_0\ll\epsilon_0$ such
that only the selected small set of charge fluctuations contribute.
The sign of the difference in Eq.~(\ref{eq:Dims}) is determined by
$\sigma$, a result which is intuitively clear since the sum over the
couplings will be greater for the spin-up transition ($\sigma=1$) in
the parallel case and for the spin-down transition in the
anti-parallel one ($\sigma=-1$), as shown schematically in
Fig.~\ref{tmr_spin}(a).
For zero energy splitting $E^{(\bar\sigma)}_{(\sigma)}$ we would
expect a broadening of the peak associated with the transition
$0\rightleftarrows (\uparrow)$ for the parallel configuration and a
broadening of the peak in $G^{\mathrm {ap}}$ for the transition
$0\rightleftarrows (\downarrow)$. Note, however, that the second
effect will not be visible since the TMR ratio will be dominated by
the spin up transition. Hence, we will observe a TMR signal as
depicted in Fig.~\ref{tmr_spin}b).

\begin{figure} \centering
\includegraphics[width=.48\textwidth]{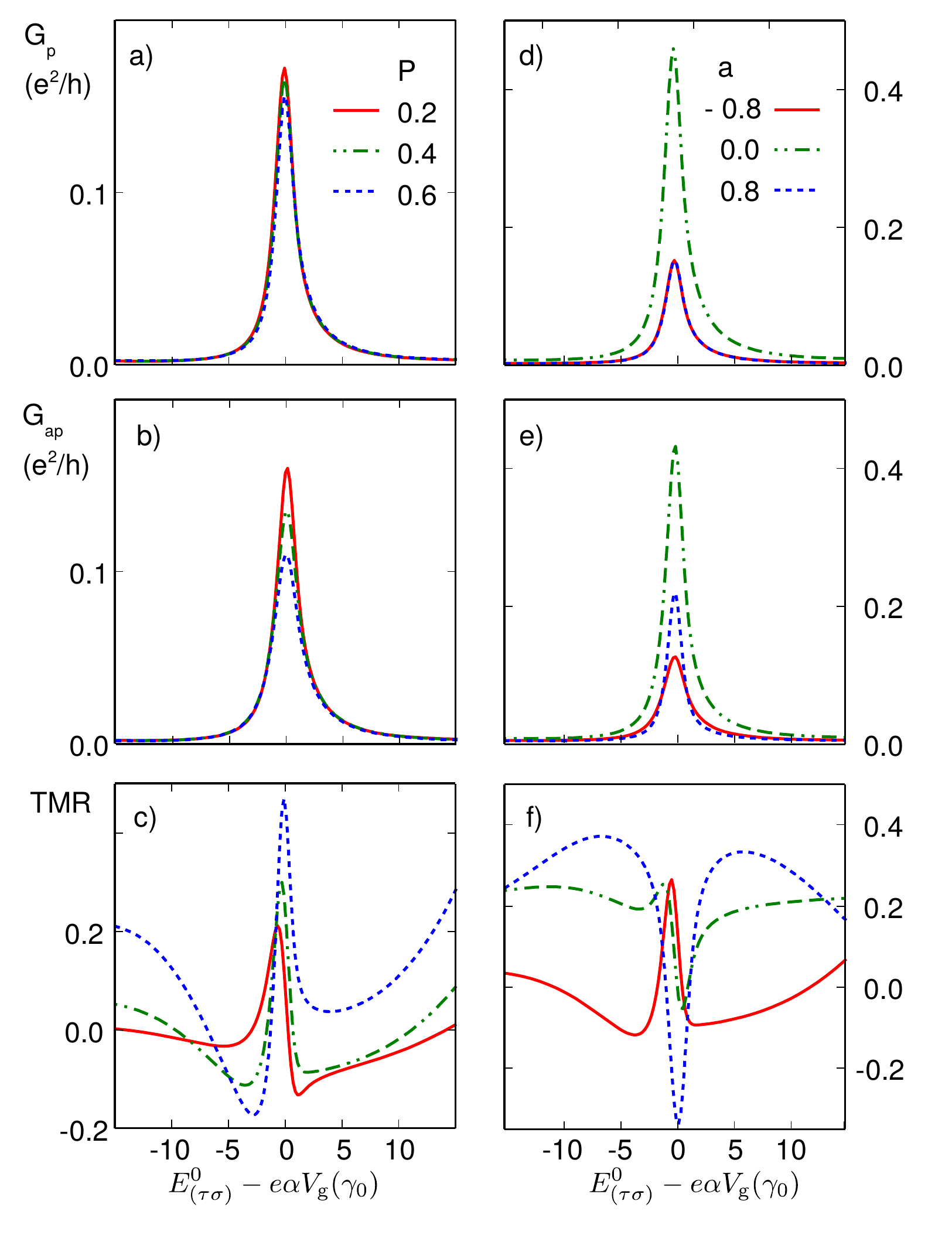}
\caption{(Color online) Conductance and TMR calculations in the
vicinity of the resonance $|0,n+1\rangle \rightleftarrows
\{|(\sigma\tau),n+1\rangle\}$ for different polarizations $P$ (panels
(a)-(c), $a=-0.8$) and coupling asymmetry $a$ (panels (d)-(f),
$P=0.4$) applied in the parallel configuration for effective Zeeman
splitting $\hp = -40$\unit{$\mu$eV} and $\hap =
-80$\unit{$\mu$eV}. (a),(b): increasing the polarization reduces the
peak width and height of both $\Gp$ and $\Gap$. (c): In the TMR curve,
the shoulder on the left at $P=0.2$ is shifted to the right for
$P=0.6$. (d),(e): The coupling asymmetry $a\neq 0$ diminishes the peak
heights of the conductance for both configurations of the leads. Note
that in the anti-parallel case shown in (e) the symmetry between the
contacts is broken and the peak height is sensitive to the variation
of the dominating coupling. (f): The TMR curve exhibits a double dip
feature for values $-1\lesssim a< 0$. It is transformed to a double
peak for $0<a\lesssim 1$. All plots are calculated at a temperature
corresponding to
$40$\unit{$\mu$eV} and a coupling
$\gamma_0=160$\unit{$\mu$eV}.\label{tmr_sim_poltau}}
\end{figure}

Now let us assume a non-zero effective Zeeman splitting
$E^\uparrow_\downarrow=E_\uparrow-E_\downarrow=g\!\mub h_{\mathrm
{p/ap}}$ of states with quantum numbers
$\sigma=\uparrow\!/\!\downarrow$. 
This splitting also depends on the
magnetization state p (parallel) or ap (anti-parallel) of the contact
electrodes. The energy difference is expressed in terms of the
effective magnetic fields $\hp$ and $\hap$. We assume that this field
is non-zero for both polarizations. $\Imp(\Sigma)$ as well as the TMR
are very sensitive to the choice of the shifts, the couplings and the
polarization. The mechanism we want to discuss can be observed for
different parameter regimes, but for the sake of the argument it is
sufficient to present one possible set that we deduce from the
experiment and the line of reasoning that goes with it. In the last
part of Sec.~\ref{sec:exc_state_ren} we argue that couplings
$\kappa_{\mathrm s}<\kappa_{\mathrm d}$, or, similarly, $0>a>-1 $ are
needed to explain the shift of the excited state lines in
Fig.~\ref{cd}. Furthermore we point out that the peaks in conductance
in Fig.~\ref{tmr3dia} are descending in height as we fill the
shell. In our model the drain lead switches polarization upon
interaction with external magnetic field while the density of states
in the weakly coupled source contact remains unaltered. Given that the
spin transport is more sensitive to the bottleneck (source) contact,
it is plausible to assume that the shifts are such that the majority
spins tunnel first on the quantum dot, namely spin up electrons in
both configurations. These considerations favor a choice of negative
shifts $\hap,\hp<-\kbt$. The second pair of resonances is then
dominated by spin down electrons and the respective contributions
$f^-(\epsilon+g\mub h_{\mathrm {p/ap}})$ in Eq.~(\ref{eq:Imdif}) are
suppressed. Conversely, for spin up electrons $f^-(\epsilon-g\mub
h_{\mathrm {p/ap}})=1$.  In the resonant case,
$|\epsilon|\lesssim \kbt$, the imaginary part of the self-energy for
the $|0\rangle\leftrightarrows|(\sigma)\rangle$ then reads
\begin{align}\label{eq:dimpap} 
\delta^{\Imp}\simeq -\pi\frac
{\gamma_0}4 (1-a)P(1-\sigma f^-(\epsilon)).
\end{align} 
The magnitude of the relative broadening of the peak
related to the transition of a spin down electron in $\Gap$ is thus
increased for higher polarization and $a\to -1$. Although this
estimate is only valid in the direct vicinity of the resonance, it
describes the situation qualitatively as can be seen in
Fig.~\ref{tmr_sim_poltau}. We show conductance and TMR nearby the
resonance $|0,n+1\rangle \rightleftarrows
\{|(\sigma\tau),n+1\rangle\}$ for fields $\hp = -40$\unit{$\mu$eV} and
$\hap= -80$\unit{$\mu$eV}. In the panels on the left side, the
polarization is varied keeping $a=-0.8$ fixed. We see that the right
shoulder in the TMR curve (c) is lifted upwards with increasing
polarization. On the right panels in Fig.~\ref{tmr_sim_poltau} we
increase the coupling to the source contact which is proportional to
$a$. While the conductance is decreased for asymmetric choices of $a$
in both configurations (see (d) and (e)), the magnitude of the peak in
$\Gap$ is not symmetric with respect to the coupling to source and
drain. The TMR in Fig.~\ref{tmr_sim_poltau}(c) can be related to
Eq.~(\ref{eq:dimpap}): the shoulders for $a=0.8$ turn into dips
approaching $a=-0.8$.
Please keep in mind that this discussion is simplified
since we do not account for the fact that the relative position of the
peaks changes, too, as we vary the parameters $a$ and $P$ (compare
$\Rep(\Sigma)$ and $\Imp(\Sigma)$ plotted in Fig.~\ref{se_ex} in the
appendix Sec.~\ref{sec:cfb}).

\section{Comparison}
\label{sec:comp}

\paragraph{Conductance in the experiment and in the model} 

\begin{figure*}
\centering
\includegraphics[width=\textwidth]{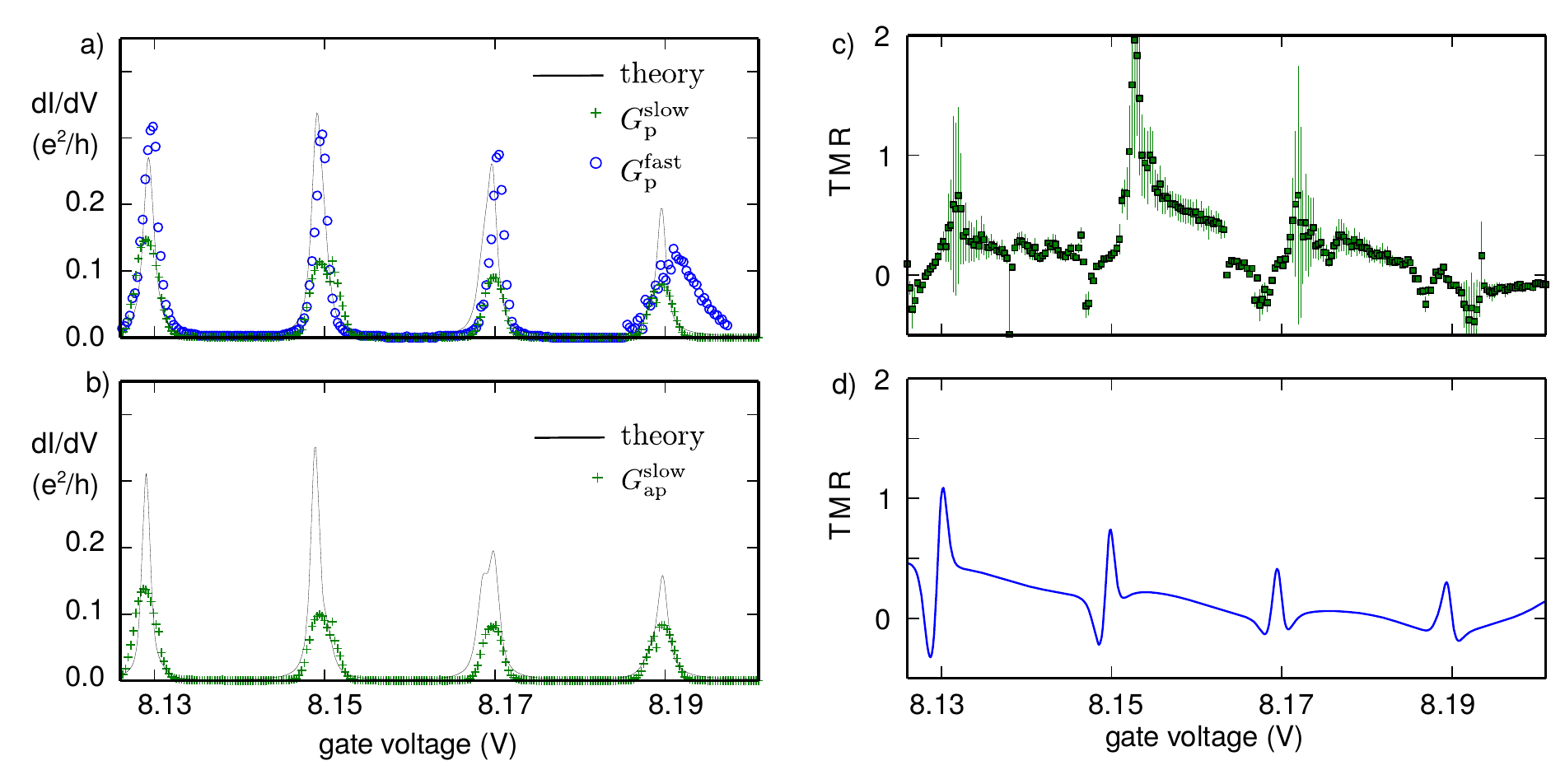}
\caption{(Color online) Conductance at zero bias as a function of gate
voltage $\Vg$ plotted for (a) parallel and (b) anti-parallel
polarization of the leads. In (a), a gate trace ($G^{\mathrm
{fast}}_{\mathrm {p}}(\Vg)$, blue circles) is shown together with
conductance obtained during TMR measurements $G^{\mathrm
{slow}}_{\mathrm p}(B,\Vg)$ (green crosses, see also
Fig.~\ref{tmr3dia}), and the calculated conductance for parallel lead
polarization (continuous line, black) at $\kbt = 40$\unit{$\mu$eV},
$\epsilon_0=1.4$\unit{meV}, $E_c= 6.1$\unit{meV}, $a=-0.7$ ,
$P=0.4$, $\hap=-0.16$\unit{meV} and $\hp=-0.12$\unit{meV}. In
the vicinity of the rightmost resonance, $G^{\mathrm {fast}}_{\mathrm
p}$ shows a high noise level (compare also Fig.~\ref{cd}). (b) The
conductance data measured for anti-parallel polarization of the
contacts $G^{\mathrm {slow}}_{\mathrm {ap}}(B,\Vg)$ (green crosses) is
compared to the model output (continuous line, black) for the same
parameters as in (a). (c) Experimental TMR data calculated from
$G^{\mathrm {slow}}_{\mathrm {p}}$(a) and $G^{\mathrm {slow}}_{\mathrm
{ap}}$(b) (also shown in Fig.~\ref{tmr3dia}). (d) TMR obtained from
the model conductance (continuous lines in (a) and
(b)).\label{tmr_gp_simexp}}
\end{figure*}

In Fig.~\ref{tmr_gp_simexp}(a) (blue circles) we show the conductance
$G_{\mathrm {p}}^{\mathrm {fast}}$ obtained at $B=0$ performing a
\textit{fast} measurement, i.e., sweeping the
gate voltage $\Vg$ at zero bias voltage, see Sec.~\ref{sec:meas}. 
Note that it provides only conductance data for the
parallel configuration (compare Fig.~\ref{fig:switch}).
The data from this measurement yields
conductance peaks that fit to Lorentzian curves with an average
$\mathrm{FWHM}$ of $0.3$\unit{meV}. Adapting our model parameters to
the data of $G_{\mathrm {p}}^{\mathrm {fast}}$, we obtain the
continuous lines in Fig.~\ref{tmr_gp_simexp}(a,b,d). The
  conductance data from the \textit{slow} measurement 
(compare Sec.~\ref{sec:meas}) for the two configurations, $G_{\mathrm
{p}}^{\mathrm {slow}}$ and $G_{\mathrm {ap}}^{\mathrm {slow}}$, are
shown in Fig.~\ref{tmr_gp_simexp}(a,b) (green crosses).
 The shape of the conductance peaks 
turns out to be non-Lorentzian, with the peak height in
the conductance data limited to $\sim 0.1$\unit{e$^2$/h}. While the
flanks of the peaks match for the first three resonances in the data
from the slow and from the fast measurement \footnote{There is a
deviation between $G_{\mathrm {p}}^{\mathrm {fast}}$ and $G_{\mathrm
{p}}^{\mathrm {fast}}$ in Fig.~\ref{tmr_gp_simexp}(a) at the right
flank of the second resonance at $\Vg = 8.15$\unit{V} due to jump in
the gate voltage during the measurement.}, the maximum conductance
values deviate by a factor of three. So far no full explanation for
the suppression of the peak conductance was found.

\paragraph{Model parameters} A bare coupling of
$\gamma_0=80$\unit{$\mu$eV} is found to optimize the fit to
$G_{\mathrm {p}}^{\mathrm {fast}}$. The thermal energy is chosen as
$k_{\mathrm B}T = 40$\unit{$\mu$eV} ($460$\unit{mK}), close to the base
temperature ($300$\unit{mK}). For the quantum dot parameters we set
$E_c= 6.1$\unit{meV} and a shell spacing
$\epsilon_0=1.4$\unit{meV} as inferred from Sec.~\ref{sec:exp}. The
shell number $n\sim 40$ is estimated from the distance to the
band-gap. We assume asymmetric contacts with $a=-0.7$ and polarization
$P=0.4$. For the calculation of the charge fluctuations
we include all states within an energy interval of $3\epsilon_0$ (see
Sec.~\ref{sec:cfb} in the appendix). 
The effective Zeeman shifts for the model output in
Fig.~\ref{tmr_gp_simexp} are $\hp=-0.12$\unit{meV} and
$\hap=-0.16$\unit{meV}. 

\paragraph{Discussion} 
If only features of the leads density of states at the Fermi
  energy are included, compare Eq.~(\ref{eq:polarization}), the 
DSO preserves particle-hole symmetry by
construction~\cite{Kern2013}. To break this symmetry, a Stoner-shift
of the majority band with respect to the minority band should be
included~\cite{Gaass2011}, whose effect is analogous to that of an
effective Zeeman field~\cite{Koller2012}. Such effective fields have
also been used to model the effects of coherent reflections at the
magnetic interfaces in double barrier systems~\cite{Sahoo2005}. 
Since the data in Fig.~\ref{tmr_gp_simexp}(a-c) does not
reflect particle hole symmetry, we use effective Zeeman splittings 
to break the particle-hole symmetry and reproduce 
the observed magnitude of the TMR effect.
The splittings are of similar magnitude as those used in Ref.~\citenum{Cottet2006}
($\hp=0.25$\unit{meV} and $\hap=0.05$\unit{meV}) to explain the
experimental TMR data of Ref.~\citenum{Sahoo2005}.

In case of non-zero spin-orbit coupling~\cite{Kuemmeth2008,Steele2013}, 
we would expect a splitting of the excited state lines in
the stability diagram in Fig.~\ref{cd}. This is not resolved 
in our experimental data. For simplicity we therefore here assume $\DSO=0$. 
Model calculations with
non-zero spin orbit coupling can be found in the appendix,
Sec.~\ref{sec:so}.

From the conductance traces calculated within
our model, Fig.~\ref{tmr_gp_simexp}(a,b) (continuous lines), the TMR,
Fig.~\ref{tmr_gp_simexp}(d), is obtained. The data and the model
calculation agree in the decay of the TMR amplitude within a sequence
of four charging states including the ``double dip'' feature in the
last two resonances at $\Vg = 8.17$\unit{V} and $\Vg =
8.19$\unit{V}. This indicates that the sequence in
Fig.~\ref{tmr_gp_simexp} represents one shell, i.e., charging states
$4n+1$ to $4(n+1)$. We note that in the model output the last
resonance is dominated by a peak while the dips are more prominent in
the experimental data. 

In the vicinity of all conductance peaks (at
$\Vg = 8.13$\unit{V}, $\Vg = 8.15$\unit{V}, $\Vg = 8.17$\unit{V} and
$\Vg = 8.19$\unit{V}) an additional small shoulder around
$\mathrm{TMR} = 0$ occurs in the data of Fig.~\ref{tmr_gp_simexp}(c). 
These shoulders are likely related to the
aforementioned suppression of the peak conductance in the \textit{slow} measurement
(see Fig.~\ref{tmr_gp_simexp}(a,b)). We recall that the TMR is
calculated from the ratio $\Gp/\Gap$ (compare also
Fig.~\ref{fig:intro} and Fig.~\ref{tmr_spin}): in the regions where
the peaks are cut off, the ratio $G_{\mathrm {p}}^{\mathrm
{slow}}/G_{\mathrm {ap}}^{\mathrm {slow}}$ is smaller than it is in
the same region in the model output, where steep peak flanks lead to a
larger ratio $G_{\mathrm {p}}/G_{\mathrm {ap}}$. 

\section{Summary}
\label{sec:sum}

The tunneling magneto-resistance of a carbon-nanotube based
quantum dot with ferromagnetic leads has been explored both
experimentally and theoretically. The experimental data shows a
distinct variation of the tunneling magneto-resistance (TMR) lineshapes
within a single quadruplet of charging states. 

To model the data we apply the dressed second-order (DSO) framework 
based on the reduced density matrix formalism. 
This theory accounts for charge fluctuations between the quantum dot
and the ferromagnetic contacts. Thereby, it goes beyond the sequential 
tunneling approximation which can only account for a positive and 
gate-independent TMR. 
 When the charge fluctuation processes are summed to all orders in the
 coupling to the leads according to the DSO
 scheme, they yield tunneling rates where the Lamb shift and the
 broadening of the resonances are given by the 
real and imaginary parts of the self energy, respectively.
This is a nontrivial result which yields the tunneling rates for an 
interacting quantum dot in the intermediate parameter regime $E_c \gg
\kbt \sim \Gamma$ depending on the polarization of the contacts.

We explicitly compare the DSO self energy for different contact
magnetizations and show that the DSO modeling can account both for the
renormalization of excited states and the specific structures observed
in the TMR gate dependence. 
A comparison of the TMR obtained from the model and from the
experimental data shows a qualitative agreement.

\begin{acknowledgments} We gratefully acknowledge discussions with
Johannes Kern, Davide Mantelli and Daniel Schmid. This work was funded
by the Deutsche Forschungsgemeinschaft (DFG) via GRK 1570, SFB 689 and
Emmy Noether project Hu 1808-1.
\end{acknowledgments}

\appendix

\section{Contribution of other excited states to the 
renormalization of the self energy}
\label{sec:cfb}

\begin{figure}
\centering
\includegraphics[width=.5\textwidth]{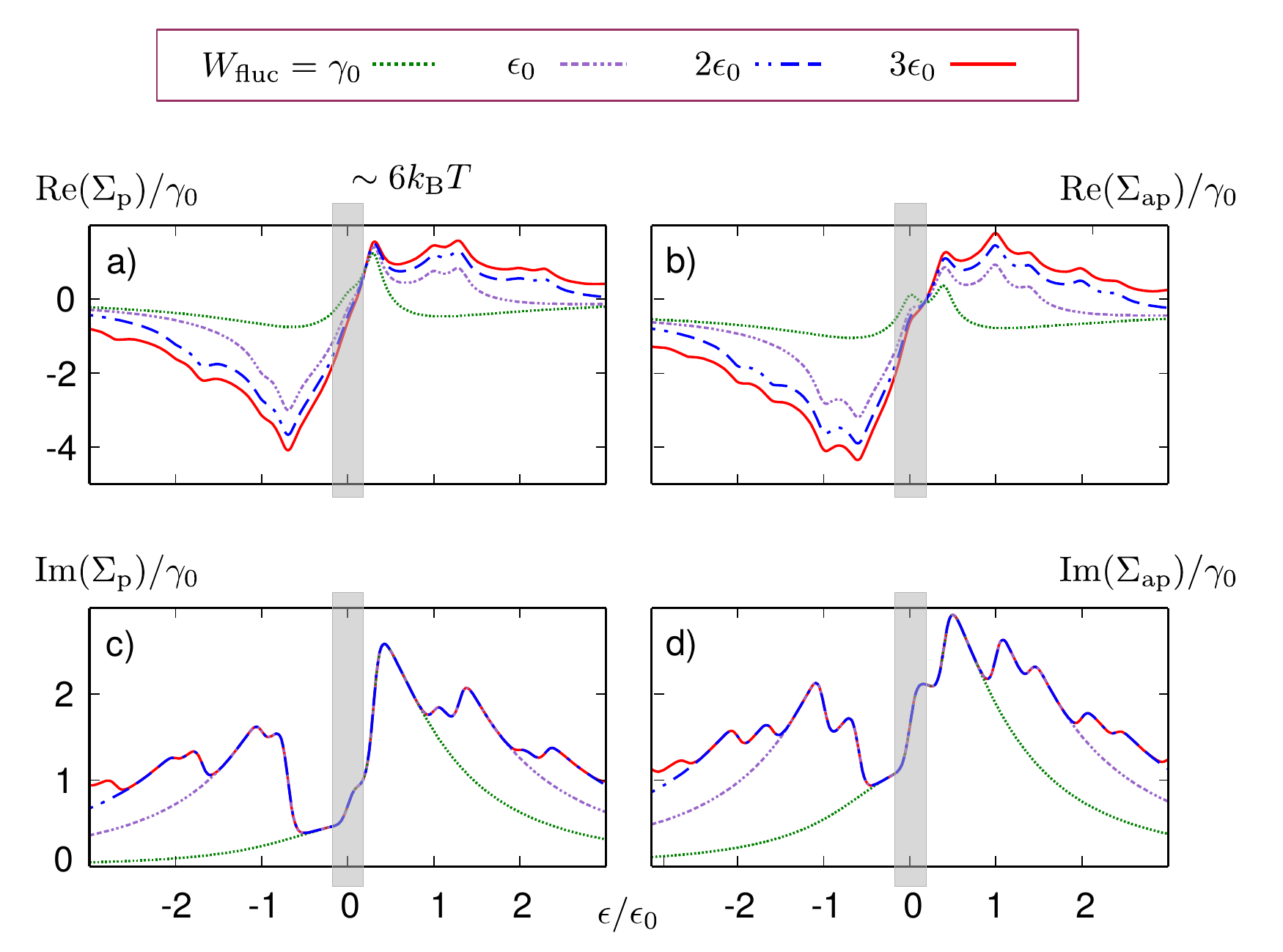}
\caption{(Color online) $\Rep(\Sigma)$ (a,b) and $\Imp(\Sigma)$ (c,d)
for both lead configurations as a function of energy $\epsilon$ in
units of the shell spacing $\epsilon_0$. Different lines are plotted
for bandwidth $W_{\mathrm {fluc}}=\gamma_0$ (green, dotted) to
$3\epsilon_0$ (red, continuous) in steps of $\epsilon_0$. In the
vicinity of a few $\kbt$ around the resonance ($\epsilon = 0$, gray
region) the difference between the graphs for the real part (a,b) is
small and for the imaginary part (c,d), it is vanishing.\label{se_ex}}
\end{figure}

When we discuss the effect of the charge fluctuations in
Sec.~\ref{sec:exc_state_ren} and Sec.~\ref{sec:tmr} of the main text,
we always focus on the most resonant transitions (see
Fig.~\ref{tmr_argument}) that are energetically favorable, i.e., on
transitions in Eq.~(\ref{eq:as}) with an energy difference
$E^{c'/b}_{a/c'}$ of the order of the effective line-width or
below. At zero bias this is the largest available energy scale in the
system. Nevertheless it is interesting to see how the outcome is
affected by increasing the bandwidth and allowing excited states of
the neighboring shells to contribute to the charge fluctuation
channels. In terms of an effective energy shift in a multi-level
quantum dot the renormalization due to excited states was also
discussed in Ref.~\citenum{Koller2012}. To illustrate the effect of
such a modification we plot the real and imaginary parts of the self
energy $\Sigma$ in the vicinity of the transition
$|(K\downarrow),n\rangle\rightleftarrows|\cdot,n+1\rangle$ for
different sets of charge fluctuations within energy ranges of
$\gamma_0$, $\epsilon_0$, $2\epsilon_0$ and $3\epsilon_0$ in
Fig.~\ref{se_ex}. We clearly see that the fluctuations from higher
shells manifest themselves in additional features in the curves for
$\Rep(\Sigma)$, Fig.~\ref{se_ex} (a,b), and $\Imp(\Sigma)$,
(c,d). Note, however, that the zero-bias conductance in our system is
only sensitive to a small vicinity of a few $\kbt$ around the
resonance. Within this range the high energy contributions do not
change the picture substantially.
The analysis of the imaginary part in Sec.~\ref{sec:tmr} is thus exact
at the level of the self energy since the Fermi functions in the
imaginary part suppress contributions from other shells.

\section{Calculation of $\Rep(\Sigma)$}
\label{sec:resigma}

In this section we perform the calculation of
$\Rep(\Sigma^{1'0})-\Rep(\Sigma^{10})$ as part of the quantity
$\delta_1$ introduced in Sec.~\ref{sec:exc_state_ren} of the main
text. To this extent we analyze the renormalization of the energy
difference $E^{1'}_{1}$ due to charge fluctuations to and from states
$0=|0;n\rangle$, $1=|K\su;n\rangle$ and $1'=|[K\su];n\rangle$ in more
detail. We recall that the real part of the self energy related to a
charge fluctuation to state $c'$ has the form (see. Eq.~(\ref{eq:as}))
\[ -\sum_l\gamma^{c'(b/a)}_l(\epsilon)\Psi^0_{\mathrm R}(\mu_l\pm
p(E^{c'/b}_{a/c'} -\epsilon)),
\] where we have to replace $b=1'$, $a=0$ or $b=1$ and $a=0$,
respectively. Note that the contribution $\propto\dig(W)$ in
Eq.~(\ref{eq:as}) does not appear explicitly since it cancels in the
difference of the shifts. Next, we have to find all states $c'$ that
contribute within our resonant approximation. We can immediately
discard states that can be reached by in-tunneling from $b$ and by
out-tunneling from $a$, since their energy differences
$E^{c'/b}_{a/c'}$ are of the order of the charging energy and thus
beyond our charge fluctuation bandwidth of $W_{\mathrm
e}=\max(e\Vb,\kbt,\gamma_0) = \epsilon_0/2$. We are left with states
that can be reached by in-tunneling into state $a$ and by
out-tunneling from state $b$. Let us discuss one example for the state
$1'$. There is one electron in the shell $n+1$ (denoted by the
brackets [...] in the state ket) which can tunnel out and we are left
with a state $|\cdot,n\rangle$. Actually this state is identical to
the state $0$ on the other contour, thus $E^{c'=0}_0=0$. We can now
evaluate the argument of the digamma function, i.e.,
$\mu_l-E^{0}_{0}+\epsilon$, for $\epsilon=0$. Since
$\mu_{\mathrm{s/d}}=\pm \epsilon_0/2$ and thus $|\mu_l|\leq W_{\mathrm
e}$, we have to sum over both leads. The total contribution from
fluctuations to $c'=0$ is thus
$-\gamma_0\sum_l\kappa_{l\uparrow}\Psi^0_{\mathrm
R}(\epsilon_0/2)$. The other states that can be reached by
out-tunneling, e.g., $|(K\su),[K\su],n\rangle$, yield energy
differences of at least $3/2\epsilon_0 > W_e$. Using similar arguments
we can collect all relevant contributions to the difference
$\Rep(\Sigma^{1'0})-\Rep(\Sigma^{10})$. In a graphical representation,
this can be visualized as

\begin{tikzpicture}[node distance=1.5cm]
\tikzstyle{tunnel}=[
nodes={fill=blue!20,minimum size=6mm,draw,inner sep=0pt,rounded corners,scale=.8},
row sep=1mm,
execute at begin cell=\node\bgroup,
execute at end cell=\egroup;%
]
\node at (0,7.5) {$\Rep(\Sigma^{1'0})-\Rep(\Sigma^{10})=$};
\node at (0,1.5) {$=2\bar\kappa_{\mathrm s}\Psi^0_{\mathrm R}(\epsilon_0/2)$};
\node [draw](label1) at (0,5){out from $1'$};
\node [draw](label2) at (5,8){in to $0$};
\node [draw](label3) at (0,2.5){out from $1$};
\node [draw](label4) at (5,2.5){in to $0$};
\node (a) at (0,6) {$-$};
\matrix [tunnel, right of=a] (m1)
{
$0$ &$$&$$&$$ \\
};
\draw (.25,5.75) to (2.75,6.25);
\node (b) at (3,6) {$+$};
\matrix [tunnel, right of=b] (m2)
{
$\epsilon_0$ & $\epsilon_0$ & $\epsilon_0$ & $\epsilon_0$\\
$0$ & $0$ & $0$ & $0$\\
$-\epsilon_0$ & $-\epsilon_0$ & $-\epsilon_0$ & $-\epsilon_0$\\
};
\draw (3.25,5.25) -- (5.75,6.25);
\node (c) at (0,4) {$+$};
\matrix [tunnel, right of=c](m3)
{
$0$ &$$ &$$ &$$ \\
$\epsilon_0$ & $\epsilon_0$ & $\epsilon_0$ & $\epsilon_0$\\
};
\draw (.25,4.) to (2.75,4.5);
\node (d) at (3,4) {$-$};
\matrix [tunnel, right of=d](m4)
{
$0$ & $0$ & $0$ & $0$\\
$-\epsilon_0$ & $-\epsilon_0$ & $-\epsilon_0$ & $-\epsilon_0$\\
};
\draw (3.25,3.5) -- (5.75,4.5);
\draw[->] (label1) to[bend right] (m1);
\draw[->] (label2) to[bend right] (m2);
\draw[->] (label3) to[bend right] (m3);
\draw[->] (label4) to[bend right] (m4);
\end{tikzpicture}
where one set of four boxes symbolizes one shell and we use
$E^{c'/b}_{a/c'}$ as a label. Fluctuations that cancel are crossed
out. Note that for excited states with an energy difference
$E^{c'/b}_{a/c'}=\pm\epsilon_0$ we add only the contribution from the
source(drain) contact where $|\mu_{l}-E^{c'/b}_{a/c'}|<W_{\mathrm
e}$. Similarly we find
\[
\Rep(\Sigma^{21})-\Rep(\Sigma^{21'}) = 
(1 + \bar\kappa_{\mathrm d}-\bar\kappa_\uparrow
+\bar\kappa_\downarrow)\Psi^0_{\mathrm R}(\epsilon_0/2),
\]
which leaves us with $\delta_1$ from Eq.~(\ref{eq:delta1}).
\section{Spin-orbit coupling and valley polarization}
\label{sec:so}
\begin{figure}
\includegraphics[width=0.5\textwidth]{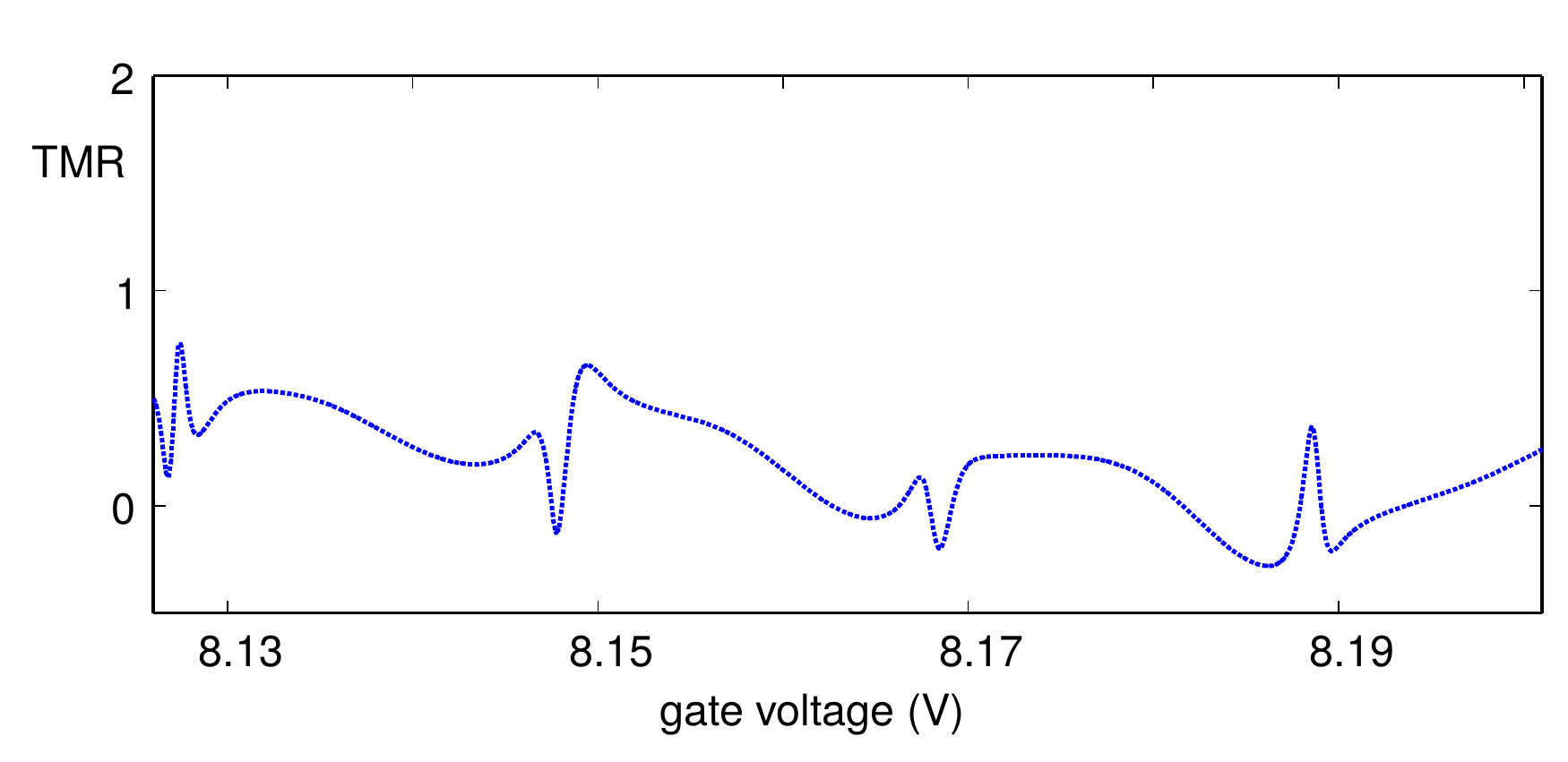}
\caption{TMR as a function of gate voltage for orbital polarization
$P_{\mathrm{orb}}=0.6$, orbital shifts $\hoap=-80$\unit{$\mu$eV} and
$\hop=-40$\unit{$\mu$eV}, and $\DSO=0.1$\unit{meV} at
$\kbt=40$\unit{$\mu$eV}. The other parameters are identical to the
ones used in Fig.~\ref{tmr_gp_simexp}.\label{tmr_so}}
\end{figure}  
In Sec.~\ref{sec:hamiltonian} we discussed the possibility to include
spin-orbit interaction effects, as they have been reported to play a
prominent role in carbon
nanotubes~\cite{Kuemmeth2008,Cleuziou2013}. However, we did not add it
in the comparison to the experimental data since they could not be
resolved in the transport spectrum (Fig.~\ref{cd}). Nevertheless,
values of the order of $\DSO\sim 100$\unit{$\mu$eV} would still be
consistent with the experimental data. Introducing a finite
$\Delta_{\mathrm {SO}}$ a priori does not affect the TMR as the
Kramers pairs are spin degenerate pairs with anti-parallel and
parallel alignment of spin and valley magnetic moments. Yet it has
been argued that the two valleys of a CNT can couple differently to
the leads~\cite{Lim2011}. If the valley quantum number is conserved
upon tunneling, the mechanism can be understood in terms of a valley
polarization. A possible tunneling Hamiltonian that describes this
situation can be written as
\begin{align}
\Ha_{\mathrm T}=\sum_{l \mathbf k n\sigma\tau}
T_{l \mathbf k n\sigma\tau}d^{\dagger}_{n\sigma\tau}c_{l \mathbf k\sigma} + \hc,
\end{align} 
with a valley dependent coupling $T_{l \mathbf k n\sigma\tau}$ and an
operator $c_{l \mathbf k \tau\sigma}$ that describes the electrons in
the leads (that are also part of the CNT). Including a valley
polarization in turn also renders the TMR sensitive to magnetic stray
fields $\hop$ and $\hoap$ along the tube axis. The orbital magnetic
moments $\morb$ are considered to be larger then $\mub$ by one order
of magnitude~\cite{Rontani2014}. In Fig.~\ref{tmr_so} we present a TMR
calculation for $\DSO=100$\unit{$\mu$eV}, orbital polarization
$P_{\mathrm {orb}}=0.6$ and stray fields $\hoap=-80$\unit{$\mu$eV} and
$\hop=-40$\unit{$\mu$eV} again combined with the experimental
data. The spin-dependent shifts are assumed to be negligible in this
setup.  We see that the agreement with the experimental data improved
slightly in Fig.~\ref{tmr_so} at the expense of additional free
parameters. It is, however, outside the scope of this paper to discuss
the effect of spin-orbit coupling and the valley polarization in more
detail.

\bibliography{bibliography.bib}

\end{document}